\documentclass[fleqn,usenatbib]{mnras}
\usepackage{times}
\usepackage{url}
\usepackage{epsfig}
\usepackage[flushleft]{threeparttable}
\usepackage{amssymb}
\usepackage{amsmath}
\usepackage{pdfpages}

\newcommand{\refbf}{}

\title[M-dwarf flares from SkyMapper]{Photometric Flaring Fraction of M dwarf Stars from the SkyMapper Southern Survey}

\author[Chang et al.]{Seo-Won Chang$^{1,2,3}$, Christian Wolf$^{1,2,3}$, Christopher A. Onken$^{1,2}$ \\
$^1$Research School of Astronomy and Astrophysics, The Australian National University, Canberra, ACT 2611, Australia \\
$^2$ARC Centre of Excellence for All-sky Astrophysics (CAASTRO), Australia \\
$^3$ARC Centre of Excellence for Gravitational Wave Discovery (OzGrav), Australia \\
}%

\begin{document}

\date{draft \today}

\maketitle

\begin{abstract}
We present our search for flares from M dwarf stars in the SkyMapper Southern Survey DR1, which covers nearly the full Southern hemisphere with six-filter sequences that are repeatedly observed in the passbands $uvgriz$. This allows us to identify bona-fide flares in single-epoch observations on timescales of less than four minutes. Using a correlation-based outlier search algorithm we find 254 flare events in the amplitude range of $\Delta u \sim 0.1$ to 5~mag. In agreement with previous work, we {\refbf observe the flaring fraction of M dwarfs to} increase from $\sim 30$ to $\sim $1\,000 per million stars for spectral types M0 to M5. We also confirm the decrease in flare fraction with larger vertical distance from the Galactic plane that is expected from declining stellar activity with age. Based on precise distances from Gaia DR2, we find a steep decline in the flare fraction from the plane to {\refbf 150~pc} vertical distance and a significant flattening towards larger distances. {\refbf We then reassess the strong type dependence in the flaring fraction with a volume-limited sample within a distance of 50~pc from the Sun: in this sample the trend disappears and we find instead a constant fraction of $\sim$1\,650 per million stars for spectral types M1 to M5. Finally,} large-amplitude flares with $\Delta i > 1$ mag are very rare with a fraction of $\sim 0.5$ per million M {\refbf dwarfs}. Hence, we expect that {\refbf M-dwarf} flares will not confuse SkyMapper's search for kilonovae from gravitational-wave events.

\end{abstract}

\begin{keywords}
stars: flare -- stars: low-mass --  stars: activity --  
stars: statistics -- techniques: photometric
\end{keywords}

\section{Introduction}\label{sec:intro}
{\refbf Time}-domain surveys are a window into the explosive universe, revealing a variety of transient sources. {\refbf The} most common high-amplitude, fast transients in {\refbf such} surveys are flares from M dwarf stars in our Milky Way. The identification of {\refbf less common} transients needs to work through this {\it foreground fog}  \citep[in the words of][]{Kulkarni2006}. {\refbf M dwarf flares} produce intense white-light continuum emission {\refbf from} near-ultraviolet {\refbf to} optical wavelengths as a generic result of magnetic reconnection, and appear with high contrast against the quiescent flux level of M dwarfs \citep{kow2013}.  This contrast increases not only with flare intensity, but also dramatically towards bluer spectral passbands, and with a decreasing surface temperature of the M dwarf. 

The All Sky Automated Survey for SuperNovae (ASAS-SN) has identified flare events, where the brightness of the dwarf star increased by up to a factor 10\,000 in $V$~band \citep{Schmidt2019}; if such events were observed in $u$ or $g$ passbands, the increase would be even larger. Hence, many flares would seem to appear, where no source {\refbf was} seen before and could not instantly be differentiated from less common types of events. {\refbf Given} the unpredictable nature of flares and their duration ranging from minutes to hours, their discovery and characterisation are limited by observing strategies in ground-based {\refbf surveys}. {\refbf In constrast, some} space-based surveys {\refbf such as the Kepler mission} offer continuous monitoring \citep[e.g.][]{Walkowicz2011, Davenport2016}. Table \ref{tab:tab1} summarises flare discoveries in several surveys. {\refbf Their} appearance is most dramatic in blue passbands, but they are {\refbf also} found at longer wavelengths \citep[e.g.][]{berger2013,chang2015,ho2018,Soraisam2018,vanRoestel2019}.

\begin{table*}
\centering
\caption{Flare discovery as fast optical transients in time-domain surveys. References: Deep Lens Survey \citep[DLS;][]{becker2004}; Sloan Digital Sky Survey Stripe 82 \citep[SDSS;][]{kow2009}; Pan-STARRS 1 Medium Deep Survey \citep[PS1;][]{berger2013}; MMT M37 Transit Survey \citep{chang2015}; intermediate Palomar Transient Factory \citep[iPTF;][]{ho2018}; All Sky Automated Survey for SuperNovae \citep[ASAS-SN;][]{Schmidt2019}; SkyMapper (this work); Kepler primary mission \citep{Davenport2016}; Transiting Exoplanet Survey Satellite \citep[TESS;][]{Gunther2019arXiv}}
\begin{tabular}{cccccccccc}
\hline \noalign{\smallskip}  
      & DLS & SDSS & PS1 & MMT & iPTF & ASAS-SN & SkyMapper & Kepler & TESS \\
\noalign{\smallskip} \hline \noalign{\smallskip}
 Telescope  & 4 & 2.5 & 1.8 & 6.5 & 1.2 & 8$\times$0.14 & 1.35 & 1.4 & 4$\times$0.1\\
 (m) &  &  &  &  &  &  & &  & \\\\
  FoV  & 2$\times$2 & 3$^\circ$ & 3$^\circ$.3 & 0.4$\times$0.4 & 7.8 & 8$\times$4.5 & 2.4$\times$2.3 & 116 & 24$\times$96 \\
  (deg$^{2}$) & & &  diameter & & & & & \\\\  
 Exposure & 600 in $BVz$ & 54 in \(riuzg\)      & 113 & 30--150 & 60 & 90 & 5--40 & 6$\times$9 & 2$\times$60 \\ 
 (sec)   & 900 in $R$   &  (drift-scan)  &   &  &  &  &  & 6$\times$270 & \\\\     
 Mag. limit & $\sim24$ & $u\sim22.0$ & $gr_\mathrm{ps1}$=23.3 & $r\sim24.7$    & $R\sim20.5$ &  $V\sim17$ & $\sim18$ &  & \\
 (mag)        &             & $g\sim22.2$ &   &     & ($g\sim21$) & ($g\sim18.5$) & (\(uvgriz\)) &  &  \\\\
 Area & High Galactic &  Stripe 82 & MDS & Open & Dec $>$ -30$^\circ$ & $\sim$16,000 deg$^{2}$ & Southern & Fixed & Two \\        
      & latitude     &  ($\sim$300 deg$^{2}$) & 10 fields & Cluster &   & (every night)  & hemisphere & field & sectors\\\\
 $\Delta t_\mathrm{interval}$   &  1000   & 108 & 16$\times$113   &  80--150  & 90   & 3$\times$90 & 240 & 60/1800 & 120 \\
 (sec)                          &           & (\(u\) to \(g\))  & (\(gr_\mathrm{ps1}\)) &    & (up to 1-day) & (one epoch) & (\(u\) to \(z\)) &  &\\\\
 $N_\mathrm{Flares}$   & 1 & 217 &  11  & 604 & 38  & 53 & 254 & 851,168 & 763 \\\\
 Reference & (a) & (b) & (c) & (d) & (e) & (f) & This work & (g) & (h) \\
\noalign{\smallskip} \hline
\end{tabular}\label{tab:tab1}
\end{table*}

The discovery of very rare, extreme flares on M dwarfs and stars of other types has reignited interest in flares in general. The Kepler mission has found evidence of superflares with {\refbf an energy of $>$10$^{34}$} erg in Sun-like stars, and we now know that single active stars can generate {\refbf such} intense events from magnetic energy stored around large starspots without help from star-planet interactions \citep[e.g.][]{Maehara2012,Notsu2013,Candelaresi2014}. In M stars such superflares might be as rare as perhaps occurring only a few times per millennium {\refbf \citep[see][]{Candelaresi2014}}, but they may impact the habitability of exoplanets and the detection of atmospheric bio-markers. By chance, two independent multi-wavelength studies discovered powerful flares from the nearest planet-hosting M dwarf, suggesting that rocky planets in that system would be inhospitable for life \citep{MacGregor2018ApJ, howard2018}. {\refbf Such} dramatic magnitude variations from flares in ultracool dwarfs (from M7 to early-L-type) provides a fresh look into the underlying dynamo action and flare generation process in the regime of low stellar mass down to the hydrogen-burning limit \cite[e.g.][]{Schmidt2014,Schmidt2016,Paudel2018ApJ...858...55P,Paudel2018ApJ...861...76P}. As new surveys opened up new discovery space {\refbf that was} found to be filled with events, we expect that future surveys {\refbf monitoring even larger volumes even more intensely} (e.g., LSST or TESS) will {\refbf push} the frontier and maybe {\refbf determine} a maximum energy for flares.

{\refbf Here, we present an analysis of M dwarf flares discovered in the SkyMapper Southern Survey \citep[SMSS;][]{wolf2018}. The survey started in 2014 to create a digital atlas of the entire Southern hemisphere in six optical passbands, and will reach a depth of $>20$~mag when completed in 2021. The SkyMapper Survey has a time-domain component providing an opportunity to detect flares, and the telescope is also used by dedicated programmes to search for extragalactic transients. In the latter context, it is relevant to know the foreground rate of M dwarf flares and its dependence on sky position, as these are the main contaminants in searches for fast extragalactic transients.}

On timescales of a day, optical afterglows from gamma-ray bursts and kilonovae associated with binary neutron star mergers are of great interest \citep{vanRoestel2019}. \citet{kow2009} measure the rate of bright M-dwarf flares across a range in Galactic latitude and longitude from observations in the equatorial SDSS Stripe 82. They find the spatial flare rate to vary with vertical distance from the Galactic disc, decreasing from 48 to 17 flares deg$^{-2}$ d$^{-1}$, but no significant trend with Galactic longitude. The Palomar Transient Factory Sky2Night programme has confirmed that M-dwarf flares are the most commonly detected Galactic transient sources, even when considering only flares that are so strong that the originator stars in their quiescent state are fainter than the survey detection limit \citep{vanRoestel2019}.

This paper is organised as follows. In Section~2 we present the time-domain data in the SkyMapper survey and the selection criteria for our M dwarf sample. In Section 3 we outline how we selected M-dwarf flares from multi-colour light curves. Section 4 highlights the properties of our flare samples and discusses the fraction of M dwarfs that flare as a function of colour and vertical distance from the Galactic plane. In Section 5 we discuss what the results imply for the search for rare transients such as kilonovae, and finally we conclude in Section 6.

\section{Data and M dwarf sample}
\label{sec:data}
We use {\refbf data release 1} (DR1) of the SkyMapper Southern Survey\footnote{\url{http://skymapper.anu.edu.au}} (\citealt{wolf2018}) to study the {\refbf flaring fraction} of M dwarfs and its dependence on colour and distance from the galactic plane. {\refbf DR1 includes} observations {\refbf from} March 2014 {\refbf to} September 2015 and {\refbf covers} $\sim $20\,000 square degrees of Southern sky. DR1 contains only observations from the SkyMapper Shallow Survey, which aims to observe sources at bright magnitudes and covers a dynamic range from magnitude 9 to 18 in all six SkyMapper filters \(uvgriz\). 

{\refbf The filters differ} slightly from {\refbf those} used by the Sloan Digital Sky Survey \citep[SDSS;][]{York2000}, and {\refbf were} designed to provide estimates of physical stellar parameters (metallicity and gravity, besides effective temperature, see \citealt{Bessell2011} for details). Note that the \(u\)~band filter (ultraviolet) is shifted toward shorter wavelength than the Balmer break at 3648\AA \ and the blue edge of the \(g\)~band is also shifted slightly redward. The \(v\)~band (violet) {\refbf lies} between the Balmer break and the Ca H\&K 4000\AA -break. Central wavelengths and widths of the other SDSS passbands vary up to 40 nm relative to the eponymous SDSS cousins.

Each visit to a given survey field consists of six short exposures in the filter sequence [\(u\)-\(v\)-\(g\)-\(r\)-\(i\)-\(z\)] (with exposure times of 40, 20, 5, 5, 10 and 20 seconds, respectively), hereafter referred to as an observation block. This sequence is enough to detect enhanced blue continuum during any stage of flare evolution (from pre- to post-maximum), because  observation blocks are executed within less than 4 minutes if there is no interruption. DR1 contains over 285 million unique astrophysical objects with more than 2.1 billion detections on individual images. A sufficiently bright source that is visible in all six filters thus accounts for six detections from each visit to their sky position. Most objects in DR1 {\refbf have} either two or three repeat visits, while in the most extreme visit numbers range from 1 to 17. The individual detections are recorded in the table \texttt{dr1.fs\textunderscore photometry}, where the multiple detections form a very sparse, randomly sampled, multi-colour light curve. 

\begin{figure}
\begin{center}
\includegraphics[width=\linewidth]{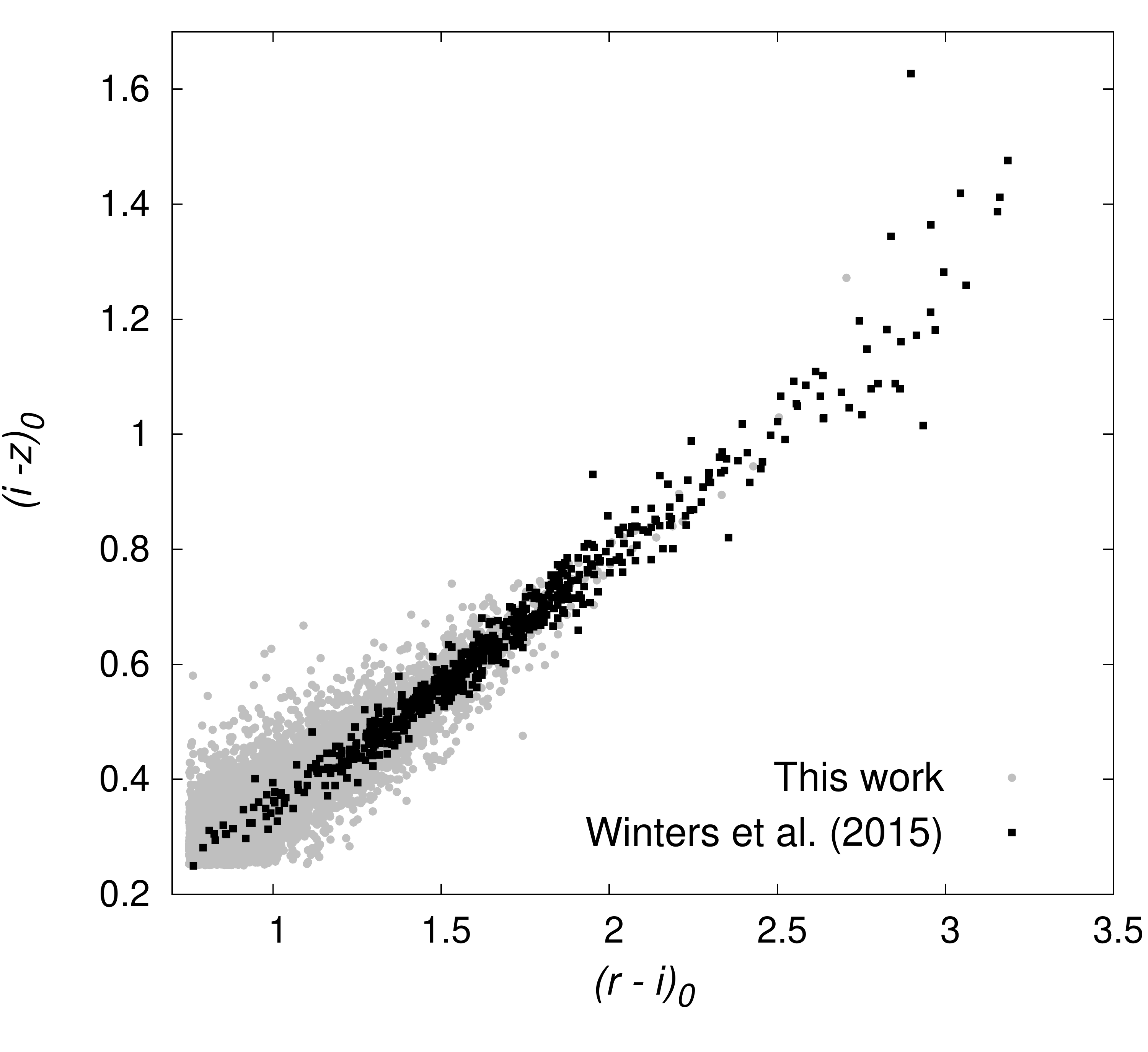} 
\caption{Colour-{\refbf colour} diagram of M-dwarfs: grey points represent the SkyMapper sample (dereddened), where we plot only 1 in every 200 objects. Black points show a cleaned sample of 622 M dwarfs (M0V--M9V) from \citet{winters2015}.}
\label{fig:SMSS Mdwarf selection}
\end{center}
\end{figure}

\begin{figure}
\begin{center}
\includegraphics[width=\linewidth]{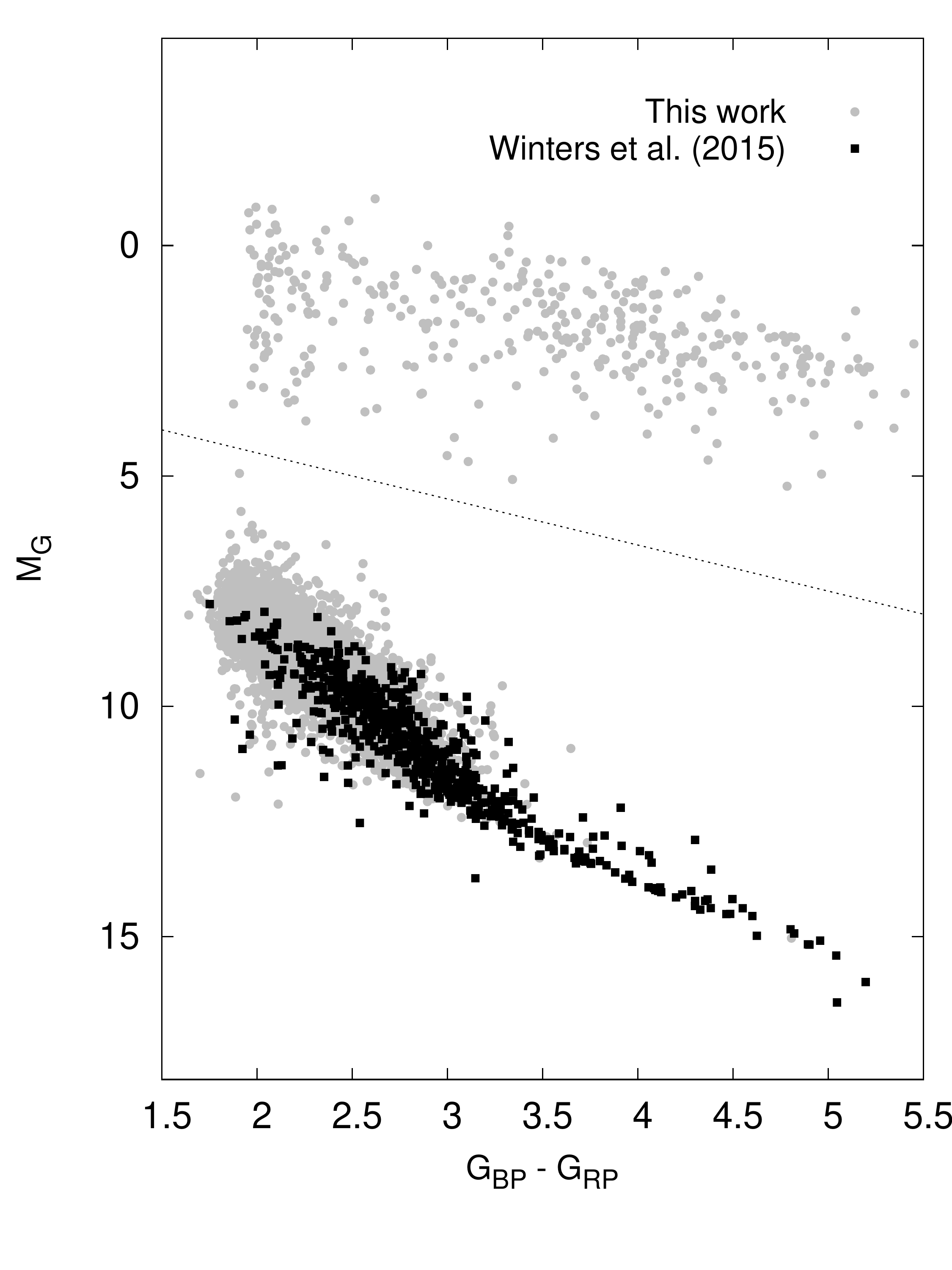}
\includegraphics[width=\linewidth]{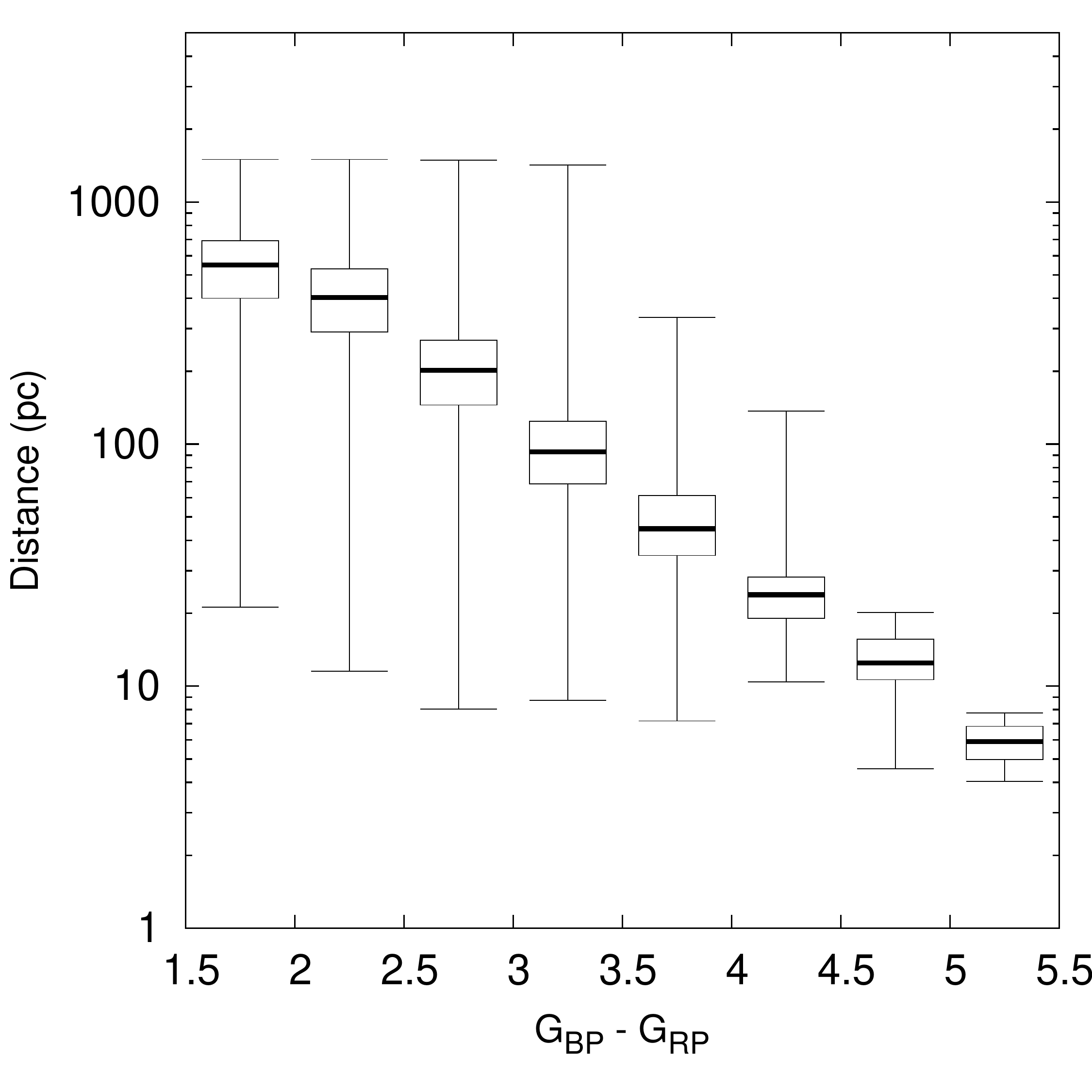}
\caption{{\refbf Top:} Colour-absolute magnitude diagram of about 1.4 million M stars with Gaia colours and parallax (symbols as in Fig.~\ref{fig:SMSS Mdwarf selection}): the dashed line clearly separates dwarfs from giants. {\refbf Bottom: Box and Whisker plot showing minimum, 1st quartile, median, 3rd quartile and maximum distances of M dwarfs for each colour bin. Note the hard cutoff at 1.5~kpc distance in our selection.
}}
\label{fig:GaiaDR2 CaMD}
\end{center}
\end{figure}

We select our sample of M-dwarfs in  three steps: first we consider the SkyMapper properties of known M-dwarfs to inform our search box; then we apply corresponding colour criteria to the full SkyMapper DR1 to create a sample of M stars; and finally we use astrometric measurements from the second data release \citep[DR2;][]{gaia_dr2_2018} of ESA's Gaia mission \citep{gaia_2016} to separate dwarfs from giants.

Our starting point is the recent compilation of Southern M dwarfs within 25pc from the Sun by \citet{winters2015}. While this catalogue does not contain all known M dwarfs in the Southern hemisphere (Dec $\leq$ 0$^\circ$), it does include 1748 genuine M-dwarfs ranging from spectral type M0V to M9.5V. We cross-match this volume-limited sample with the SkyMapper \texttt{master} table of the release version DR1.1 while taking into account proper motion: we find SkyMapper matches for $\sim97$\% of the sources, whereas the remaining 3\% are missed due to incomplete sky coverage in DR1. When comparing positions with external catalogues, we simply use the mid-time of the observations, mid-October 2014, as the reference epoch for DR1 sources. 

Next, we apply the following data quality cuts:
\begin{itemize}
    \item \texttt{FLAGS<4}: the flags are a combination of Source Extractor flags and bespoke quality flags from the SkyMapper Science Data Pipeline \citep[SDP; for details see][]{wolf2018} that warn against unreliable photometry.
    \item \texttt{prox>5}: a proximity criterion that requires the nearest neighbour source to be at least 5~arcsec away as we don't trust the photometry otherwise.
    \item \texttt{nch\textunderscore max=1}: {\refbf this means} that the source is not detected as a blend of multiple sources.
    \item \texttt{ebmv\textunderscore sfd<0.2}: moderate interstellar reddening as described by the maps of \citet{Schlegel1998}.
    \item \texttt{\{F\}\textunderscore nvisit$\geq$2}: requires {\refbf two or more} visits in each filter \{F\} {\refbf (with \{F\} in \{$uvgriz$\})}, to include the source in the list.
\end{itemize}

We do not deredden the photometry of these nearby M dwarfs as their distance places them inside the dust-poor local bubble, {\refbf which extends more than 50~pc in all directions} \citep{Frisch2011}. The sample forms a well-defined locus in the \(riz\) diagram which meets the following criteria:
\begin{equation}
    0.75 < r - i < 3.5 ~,~ 0.25 < i - z < 1.7 ~,~ r < 18 ~.
\end{equation}
It is not necessary to consider \(uvg\) colour cuts because these red objects, even at close distance, tend to be very faint in those bands.

Next, we select a sample of M-star candidates with the same criteria from the overall DR1 catalogue. Here, we correct the observed colours for reddening using the reddening coefficients listed by \citet{wolf2018}\footnote{\url{http://skymapper.anu.edu.au/filter-transformations/}}. We obtain a flux-limited sample of 1\,504\,379 objects covering nearly the full Southern hemisphere except for the higher-reddening zone at low Galactic latitudes. Fig.~\ref{fig:SMSS Mdwarf selection} shows our selection region in the $riz$ colour-colour diagram and contains both our candidate sample and the known M dwarfs (black points) from \citet{winters2015}.

As a last step, we remove M giants from our sample, which are indistinguishable with only \(riz\) colours. Hence, we restrict the sample to sources with a full five-parameter astrometric solution (position, proper motions and parallax) from Gaia DR2 to make a colour-absolute magnitude diagram (CaMD). DR1.1 has been pre-matched to Gaia DR2 and the SkyMapper data portal contains a copy of the Gaia catalogue as a separate table \texttt{ext.gaia\textunderscore dr2}.  However, this pre-match does not take proper motion into account, and hence we redo the match with proper motion folded in. Sometimes, we find ambiguous matches to more than one Gaia object, and we choose to clean the sample simply by dropping those stars where the Gaia magnitude indicates a star of the wrong brightness. We thus ensure that we get a clean sample for reliable distance estimates. We also avoid the risk of the SkyMapper magnitude being compromised by blended sources when Gaia offers more than one possible counterpart.

A simple parallax inversion gives a poor distance estimation in the presence of non-negligible errors on the parallax $\varpi$ (see \citealt{Luri2018,bailer-Jones2015} for a detailed discussion). Instead, we adopt the Bayesian approach proposed by \citet{Astraatmadja2016} to infer the most probable distance, in which we use a prior that varies with distance according to an exponentially decreasing space density distribution (with a length scale \(L\) = 1 kpc). We then remove all objects that would be further than 1.5 kpc to restrict our analysis to disc stars in the solar neighbourhood. For the majority of these sources, the relative precision on $\varpi$ is better than 20\%, which means that we {\refbf would also be able to} simply estimate the absolute Gaia magnitude using the distance modulus equation (e.g., \citealt{GaiaHDR2018}). 

Fig.~\ref{fig:GaiaDR2 CaMD} shows a clear separation between dwarfs and giants on the {\refbf CaMD. We choose to remove M giants \citep[some of which could turn out to be long-period variables, e.g.][]{mowlavi2018} by placing a final cut in luminosity within the gap, using $M_G<2.5+(G_{BP}-G_{RP})$. At this point,} our sample includes 1,386,114 objects with 4,361,814 observation blocks. {\refbf In our flux-limited sample, early- to mid-type M dwarfs ($G_\mathrm{BP}-G_\mathrm{RP} \leq 3.5$) can be seen out to distances of 1.5 kpc, while later types cover smaller volumes (see bottom panel in Fig.~\ref{fig:GaiaDR2 CaMD}). Note, that we do not de-redden $G_\mathrm{BP}-G_\mathrm{RP}$ colours because we do not have 3D extinction maps, but we checked how large the influence might be, but the most extreme cases in our sample are distant M0 dwarfs along specific lines-of-sight at low $b$. There, we find a colour correction of  $E(G_\mathrm{BP}- G_\mathrm{RP}) \la 0.25$~mag, which is small relative to the full colour range of 4~mag. We checked that the result figures would not change significantly.}

\begin{figure}
\begin{center}
\includegraphics[width=0.95\linewidth]{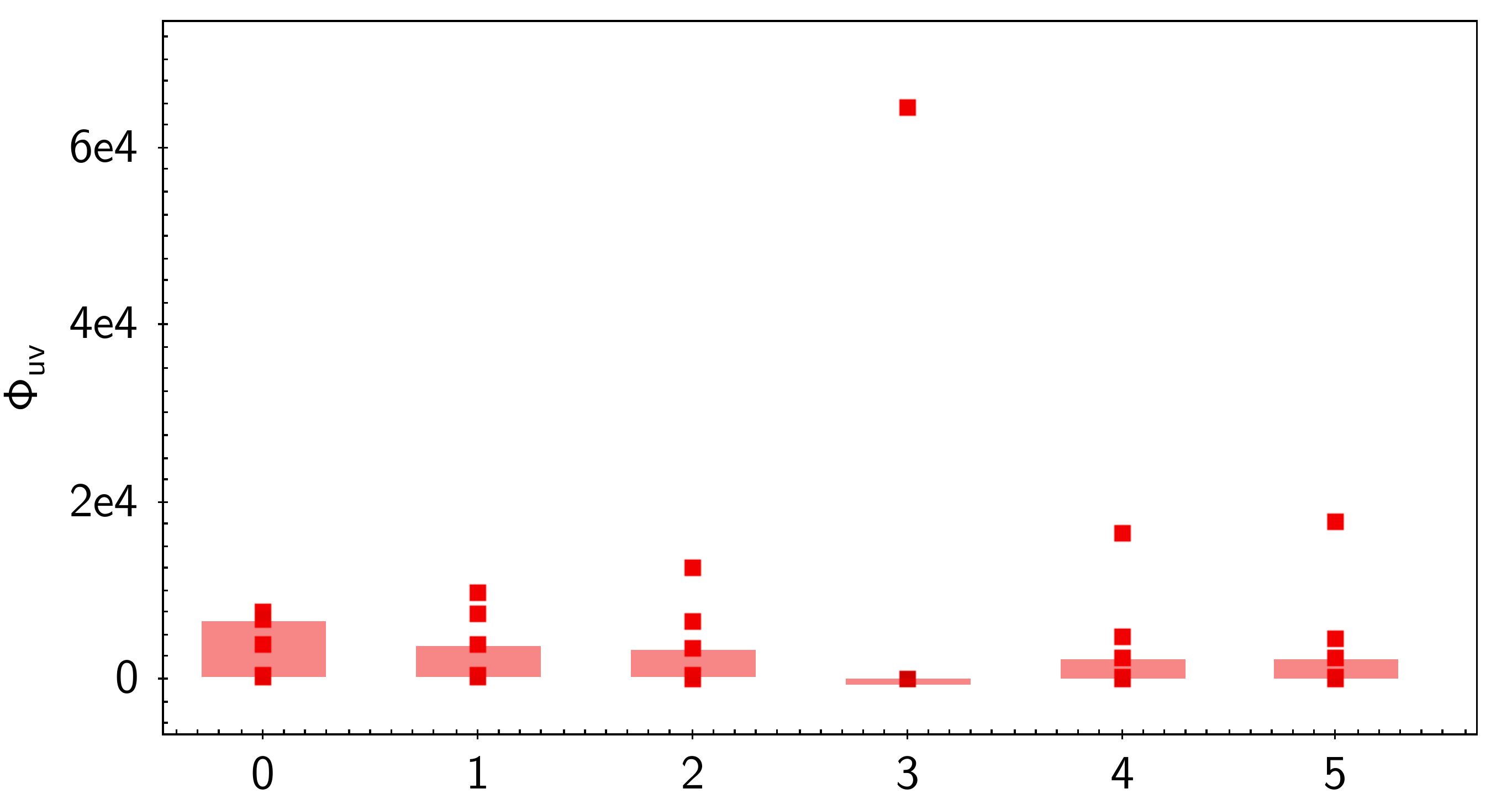}
\includegraphics[width=0.99\linewidth]{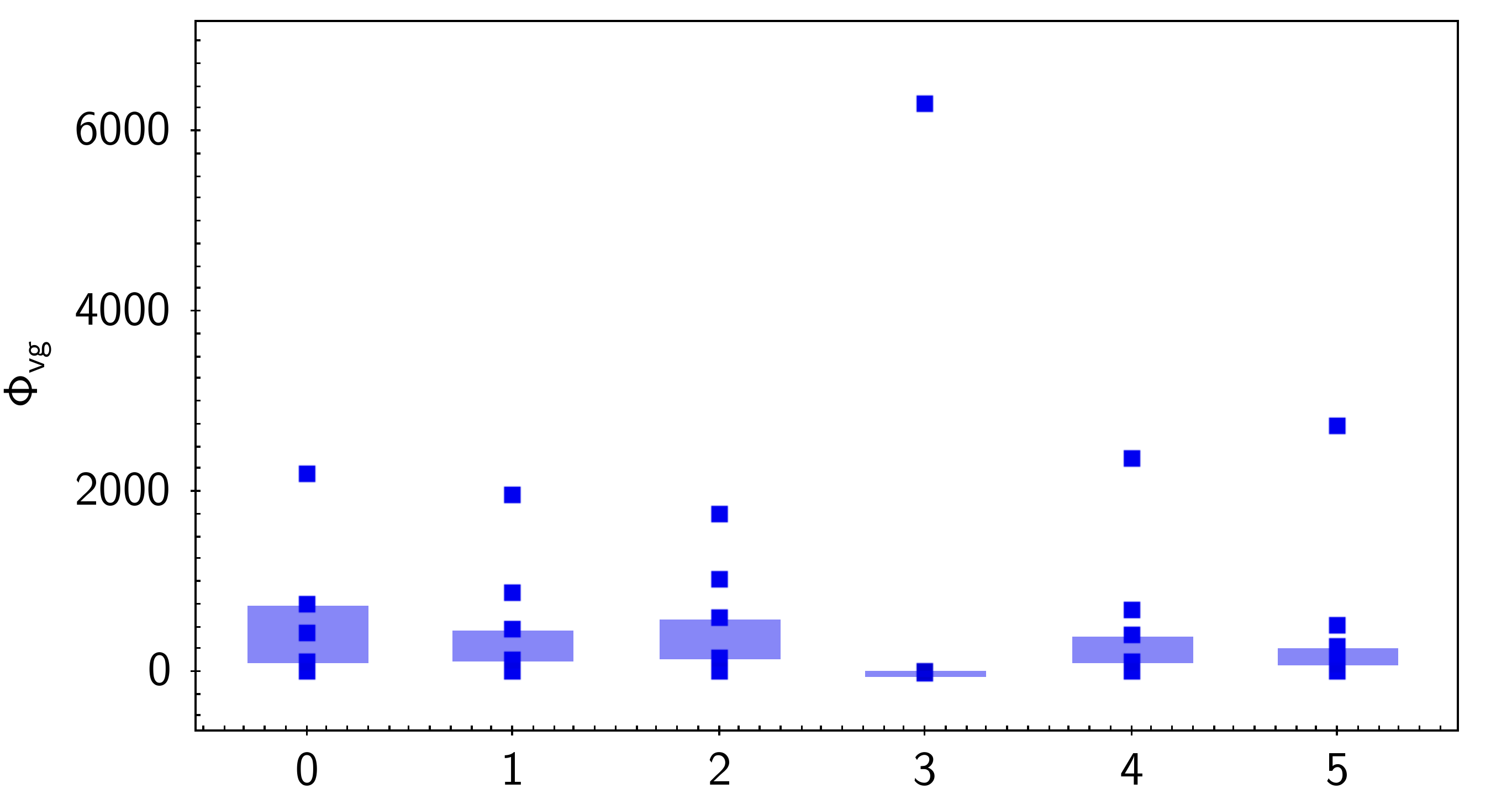}
\includegraphics[width=0.97\linewidth]{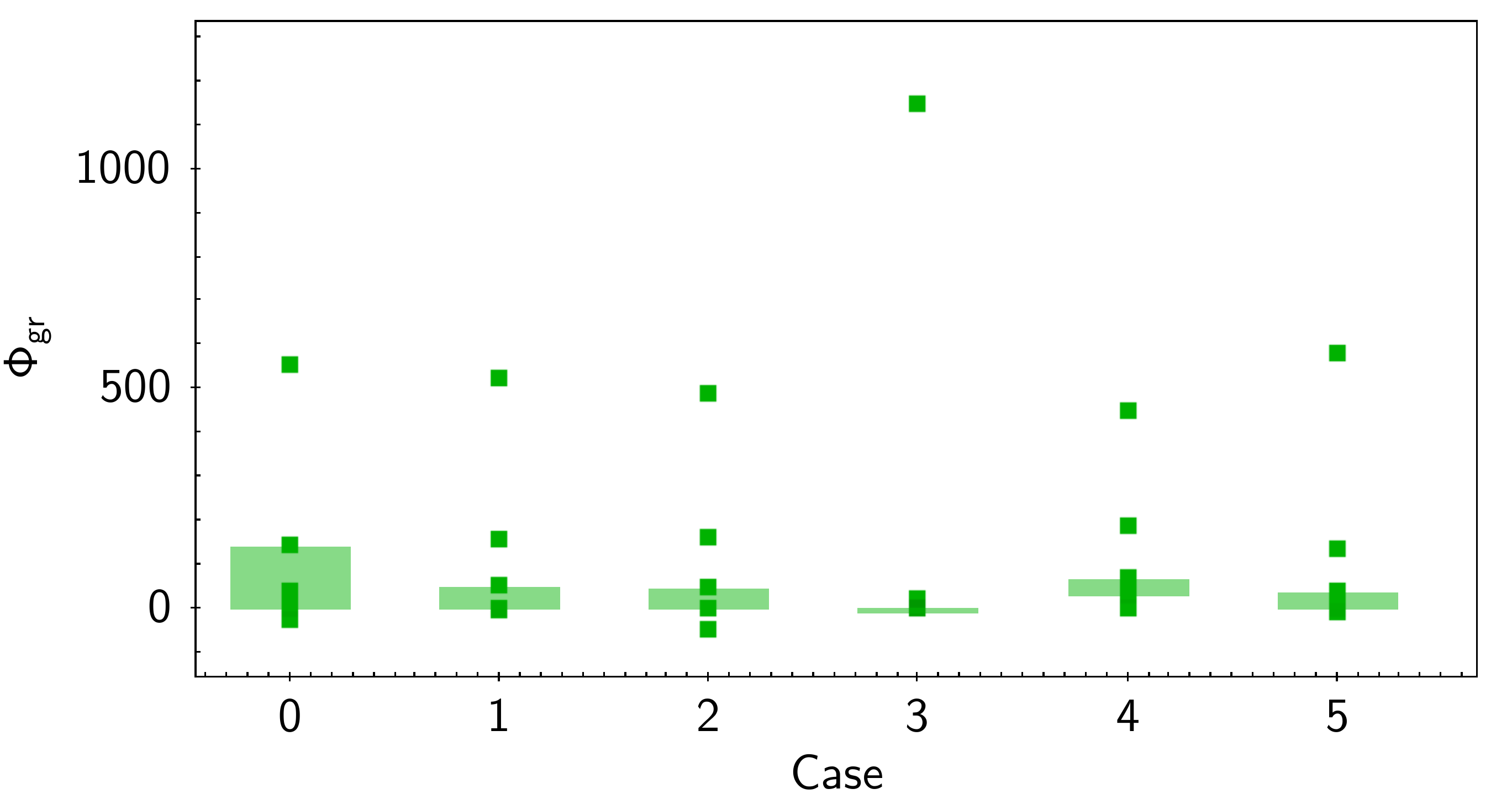}
\caption{Schematic diagram of the leave-one-out cross-validation algorithm to estimate the quiescent magnitudes of M dwarfs. The full light curve of this example object consists of five observation blocks that are indicated by a label on the x-axis (from \texttt{Case 1} to \texttt{Case 5}). \texttt{Case 0} shows an initial guess of ${\Phi}_{ij}$ values {\refbf using all data points}, while in the other cases $\hat{\Phi}_{ij,n}$ statistics are calculated by dropping the data points for epoch $n$. The shaded bars {\refbf represent} the $1 \sigma$ RMS scatter of the flare variability indices.}
\label{fig:quiescent level}
\end{center}
\end{figure}

\begin{figure}
\begin{center}
\includegraphics[width=\linewidth]{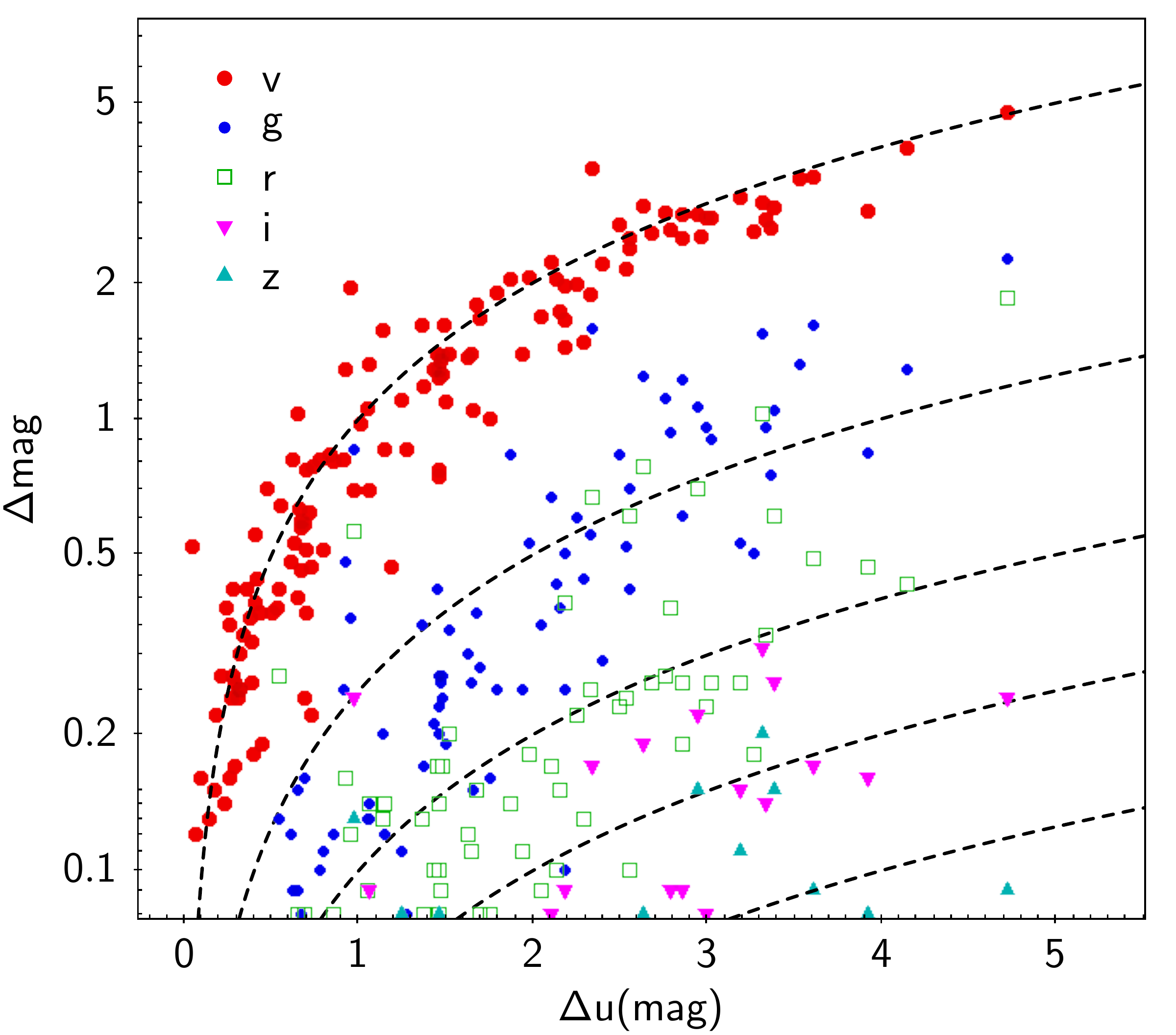}
\caption{Relation between the \(u\)~band amplitude and the near-simultaneous multi-colour (\(vgriz\)) amplitudes for 129 M-dwarf flares {\refbf detected with $\Phi_{uv}$}. Dashed lines indicate amplitudes relative to the \(u\)~band, from top to bottom: 100\%, 25\%, 10\%, 5\%, and 2.5\%, respectively.}\label{fig:flare amplitudes}
\end{center}
\end{figure}

\section{Correlation-based flare detection: dm-dt-filter pairs}
\label{sec:method}
We identify flare candidate events using a flare variability index $\Phi$ formed from filter pairs in each observation block \(n\). {\refbf This method} was also used by previous flare searches \citep[e.g.][]{kow2009, davenport2012}). The flare variability index can be written in general form as
\begin{align}
\Phi_{ij,n} =  [\frac{m_{i,n} - \overline{m}_{i}}{\sigma_{i,n}}] [\frac{m_{j,n} - \overline{m}_{j}}{\sigma_{j,n}}] ~,
\end{align} 
where \(i\) and \(j\) denote two different filters, e.g., \(uv\), \(vg\), or \(iz\). While $\overline{m}_{i,j}$ are the quiescent magnitudes of those filters, $m_{i,j}$ are single measurement points (\texttt{mag\textunderscore psf}) for each filter at epoch $n$, and $\sigma_{i,j}$ are their errors (\texttt{e\textunderscore mag\textunderscore psf}). 

The main difficulty in searching for flaring epochs is to properly estimate the quiescent flux level, which is a bit challenging as most sources in DR1 have less than five observation blocks. In sources where we observe a flare, the mean magnitudes do not represent the quiescent level but an average biased by the flare itself. {\refbf Thus, we estimate} the quiescent flux level by examining the light curve with a {\refbf rolling} leave-one-out strategy that {\refbf removes possible} flare epochs. To avoid artefacts, we only consider photometric measurements {\refbf that have} no bad pixels contained in an object's aperture, as indicated by \texttt{nimaflags=0}. {\refbf Our search for flare candidates loops over all objects in the M dwarf parent sample and considers each possible flare epoch $n$ separately; leaving out epoch $n$, we estimate the quiescent flux level and the $\hat{\Phi}$ statistic}:
\begin{align}
\hat{\Phi}_{ij,n} =  [\frac{m_{i,n} - \overline{m}_{i,n}}{\sigma_{i,n}}] [\frac{m_{j,n} - \overline{m}_{j,n}}{\sigma_{j,n}}] ~,
\end{align} 
where $\overline{m}_{i,n}$ and $\overline{m}_{j,n}$ are the weighted means obtained by dropping the measurements for epoch \(n\). In a noisy case without outliers, the flare variability index is close to a normal distribution with a mean of $\hat{\Phi}_{ij} \approx 0$. In a flaring star, the root-mean-square of the leftover measurements will shrink dramatically, when the flaring epoch is left out, clearly marking the epoch to be left out. We thus simultaneously get unbiased estimates of the quiescent magnitude and the flare variability index. 

When flares are weak and close to the noise level, we can still detect them as to be left out by combining the answer from all six passbands. Our observing sequence creates a strong correlation among the six measured magnitudes for the same epoch $n$. 

Fig.~\ref{fig:quiescent level} shows an example of the $\hat{\Phi}$ statistic from a flaring M dwarf: the initial quiescence level of $\hat{\Phi}$ in this star is overestimated due to the outlying flare measurements in the third observation block. This effect is largest in the bluest bands, where the flare emission shows the highest contrast over the quiescent flux \citep[e.g.][]{Bochanski2007, kow2013}. 

Now we select flare candidates {\refbf by looking for brightening in adjacent filters occurring nearly simultaneously within observation blocks. We require:} \texttt{$m_{i,n} - \overline{m}_{i} < 0$ AND $m_{j,n} - \overline{m}_{j} < 0$ AND} \texttt{$\Phi_{ij}$ > 0}. However, candidate signals can also be produced by correlated random noise in our light curves. \citet{Miller2001} describe the false discovery rate (FDR) approach as one way to control the number of false positives. By comparing the flare candidate distribution (\texttt{$\Phi_{ij}$ > 0}; test hypothesis) with an underlying noise distribution caused by random photometric errors (\texttt{$\Phi_{ij}$ < 0}; null hypothesis), we can set a threshold value tuned to a desired low level of false positives for each filter pair separately. We determine the probability distribution function for each null-hypothesis estimator, which is well approximated by a normal distribution, by calculating the \(p\)-values for the test distribution. By setting the FDR parameter to $\alpha=0.05$, we ensure that our flare sample ends up with {\refbf a} contamination fraction of only 5\%. This fraction corresponds to threshold values of 540, 420, 880, 755, 765 for $\Phi_{uv}$, $\Phi_{vg}$, $\Phi_{gr}$, $\Phi_{ri}$, $\Phi_{iz}$, respectively. 

\begin{figure}
\begin{center}
\includegraphics[width=0.92\linewidth]{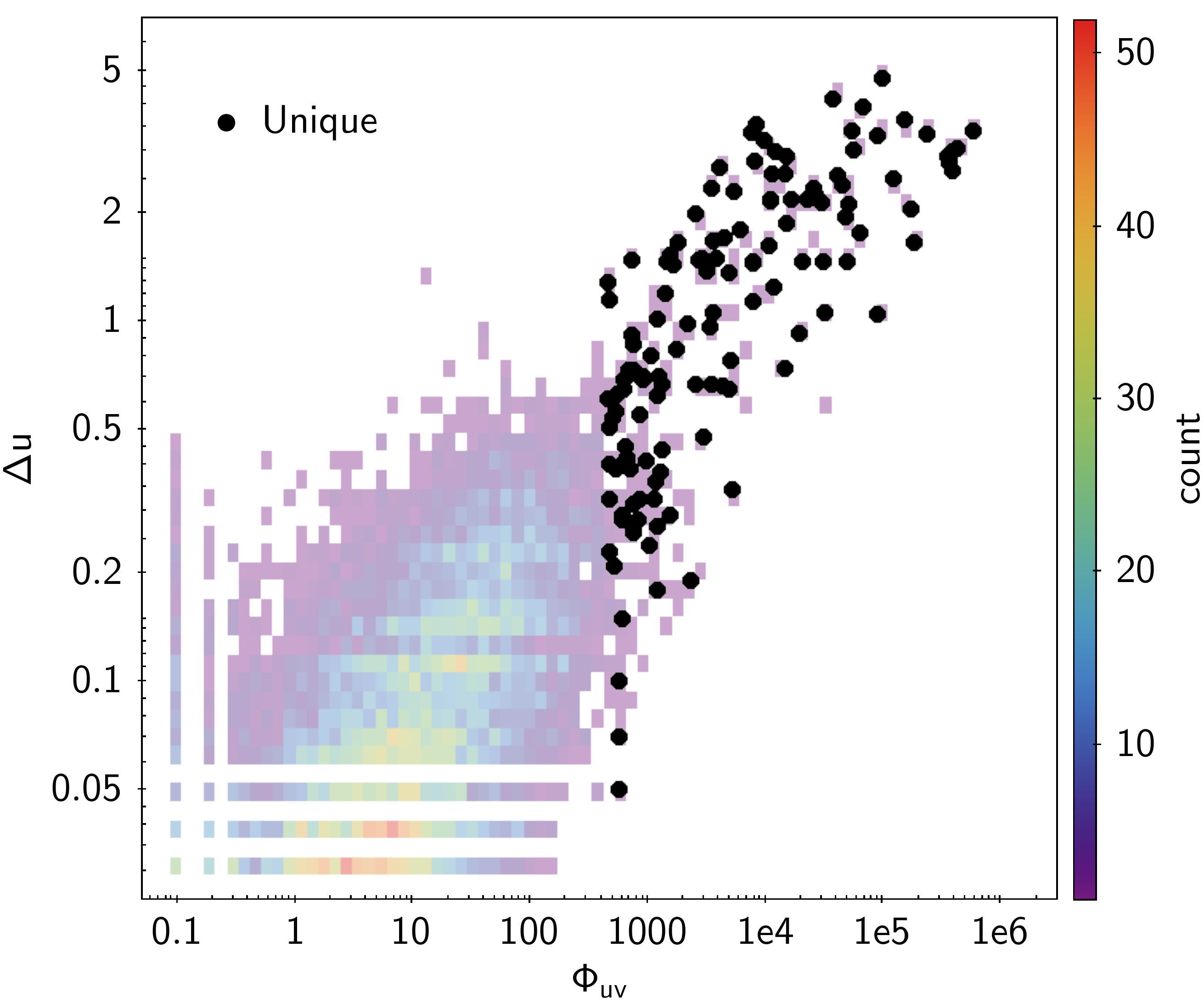}
\includegraphics[width=0.96\linewidth]{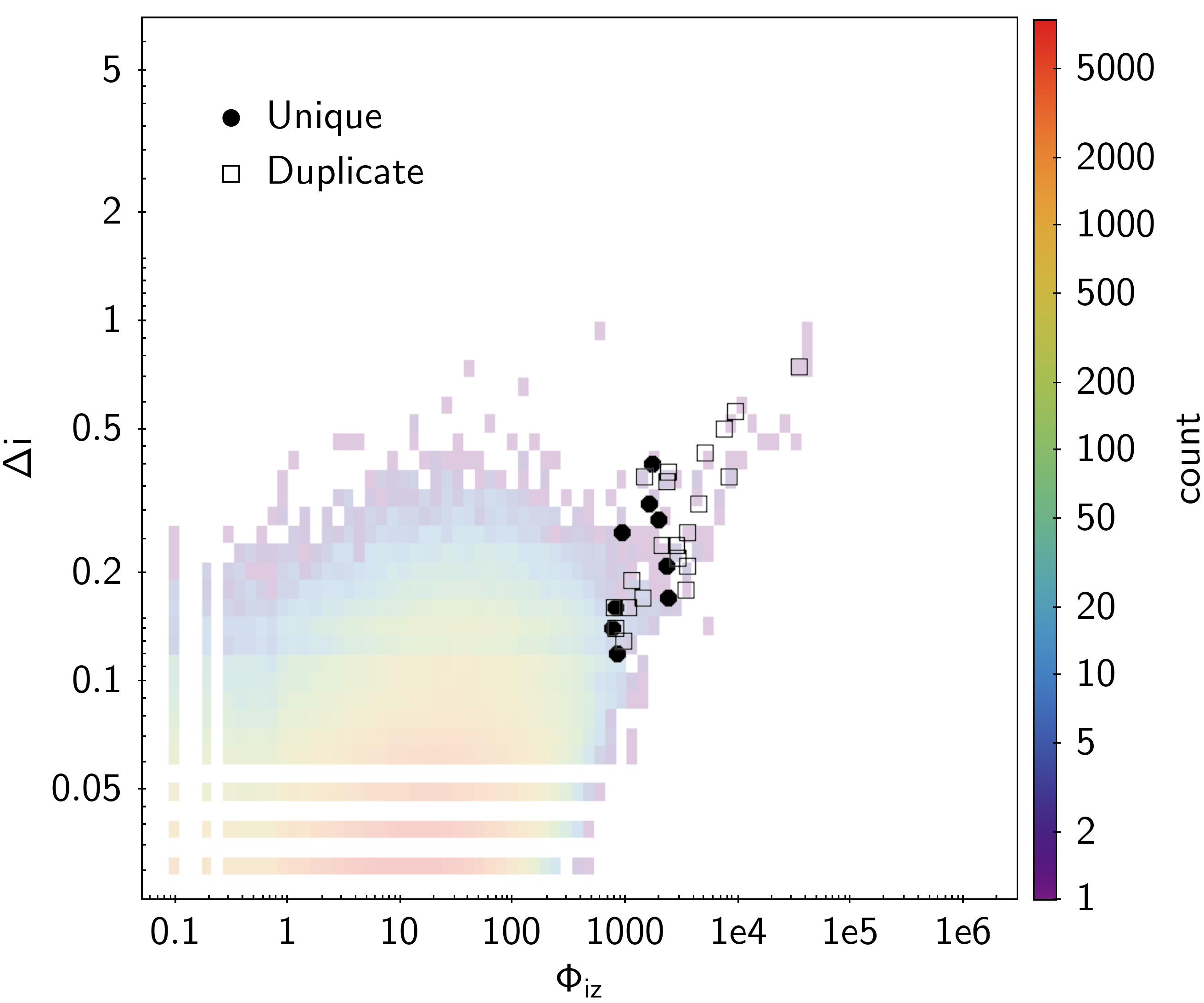}
\caption{Flare amplitude vs. flare variability index in the most (top) and least (bottom) sensitive filter pairs. The background colour maps show the number density of objects. Towards the top right, we see flares, {\refbf whether they are unique (solid symbols) or duplicate (open symbols). Duplicates are flares detected in more than one filter pair.}}
\label{fig:phi vs delmag}
\end{center}
\end{figure}

\begin{table*}
\caption{Summary of flare candidates from SkyMapper DR1}
\centering
\begin{tabular}{cccccccc}
\hline \noalign{\smallskip}  
&   & \multicolumn{6}{c}{Number of flare epochs} \\
\noalign{\smallskip} \cline{3-8} \noalign{\smallskip} 
$\Phi_{ij}$ &  Number of observed epochs & $\Phi_{ij} > 0$ & $\alpha \simeq 0.05$ & SNR $\geq 3.0$ & Eyeball & Unique & Duplicate \\
\noalign{\smallskip} \hline \noalign{\smallskip}
$\Phi_{uv}$  &     46,758 &     9,362 &  200 &  162 & 129 & 129 & -   \\
$\Phi_{vg}$  &     63,711 &    13,881 &  146 &  131 & 101 &  27 & 74 \\ 
$\Phi_{gr}$  &  2,398,357 &   566,859 &  154 &  130 & 105 &  48 & 57 \\ 
$\Phi_{ri}$  &  3,834,884 &   924,550 &  196 &  143 & 81 &  41 & 40 \\ 
$\Phi_{iz}$  &  4,075,290 & 1,015,531 &  171 &  118 &  29 &   9 & 20 \\
\noalign{\smallskip} \hline \noalign{\smallskip}
Total        & 10,419,000  & 2,530,183 &  867 & 684 & 445 & 254 & 191 \\ 
\noalign{\smallskip} \hline
\end{tabular}
\label{tab:tab2}
\end{table*}

Next, we clean the candidate sample further by removing events that are too weak to provide conclusive evidence for bona-fide flares. Fig.~\ref{fig:flare amplitudes} gives an empirical relation between \(u\)~band and \(vgriz\)~band amplitudes of flare candidates selected by the $\Phi_{uv}$ index. As expected, there is a one-to-one correlation between $\Delta u$ and $\Delta v$ but the resulting $\Delta m_{i} (=|m_{i,n} - \overline{m}_{i}|$) for the \(griz\) bands are an order of magnitude lower than the \(uv\) bands, even in large-amplitude flares of $\Delta u,v > 1$ mag. This wavelength-dependent variability is an ubiquitous characteristic of M-dwarf flares {\refbf and} reproduced by a semi-analytic two-component flare model \citep{davenport2012}. Thus, our efficiency of recovering low-amplitude flares is limited by the light curve root-mean-square ($\sigma_{LC,i}$) scatter as a function of source magnitude. In the case of the \(uv\)-filter pairs, where enhanced blue continuum is significant during flares, the trade-off is an increase in the noise level due to the faint nature of M dwarfs in blue optical bands. The flare variability is measured with a signal-to-noise ratio given by SNR = $\Delta m_{i} / \sigma_{LC,i}$. To construct a fairly complete sample of flares, we apply the same SNR threshold of 3.0 to all filter pairs. This implies the error on our measurement of the amplitude of variability can be up to $\sim$33\%. Fig.~\ref{fig:phi vs delmag} shows the relation between flare variability index and flare amplitude in the most {\refbf and least} sensitive filter pairs. Plausible candidates {\refbf are found} in the outlying low-density regions. {\refbf Some events in those regions are not marked as candidates, but these are cases where the magnitude {\it decreases} in both bands simultaneously (\texttt{$m_{i,n} - \overline{m}_{i} > 0$ AND $m_{j,n} - \overline{m}_{j} > 0$ AND} \texttt{$\Phi_{ij}$ > 0}), so these are possible eclipse-like events if they are significant.}

Finally, we eyeball the multi-colour light curves of all candidate events and remove unreliable light curves, whose (i) overall variability {\refbf does} not follow the pattern of a typical flare variation {\refbf in all six bands (i.e. a rapid rise that decreases with wavelength)} or (ii) variability signal is still dominated by systematic errors {\refbf (i.e. the observed changes in magnitude were comparable to the uncertainties for some filter pairs).}

\section{Results and Discussion}
\label{sec:results and discussion}
Using our conservative detection thresholds for searching dm-dt-filter parameter space, we find 445 flare {\refbf signals from 254 unique stars} among 10,419,000 observed epochs in any filter combinations (summarised in Table \ref{tab:tab2}). Ten of our flaring stars are listed in the AAVSO International Variable Star Index \citep[VSX;][]{Watson2006,Watson2017}, which is cross-matched to DR1 in table \texttt{ext.vsx}. They are classified as eruptive stars (INS and UV), spotted stars (ROT), and active binary systems (RS and EA).

We also use the SkyBoT cone-search service \citep{ber2006} to check the list of available solar system objects (e.g., asteroids, planets and comets) located in our field-of-view at all given epochs. We find no flare events that were cross-matched with the position of any known moving objects\footnote{We did, however, find an interesting case of an asteroid being blended with a star, when extending the SkyBoT search to red giants: a giant of 15~mag was blended with the $V\approx 10.6$~mag Inner Main Belt asteroid (230) Athamantis for the duration of one visit in June 2014, see \citealt{Onken2019} for details.}.

\subsection{Large-amplitude flares: $\Delta m > 1$ mag}
{\refbf Flares} with amplitudes larger than 1 mag can be found most easily at the blue end of the spectrum due to the higher contrast. {\refbf In} \(u\)- and \(v\)~band we find 69 and 79 large-amplitude flares, respectively, which {\refbf combine} to 92 unique flares in total. The numbers rapidly decline towards redder passbands, with 17 and 2 flares in \(g\)- and \(r\)~band, respectively. 

Even the largest \(uv\) flares in our sample {\refbf show a flux enhancement of only 0.1 to 0.3~mag} in our reddest \(iz\) filters, {\refbf where} we find not a single large-amplitude flare. This already implies that potential searches for extragalactic transients will have to deal with much less foreground fog if done in the \(iz\) bands. However, in Section~\ref{LargeFlaresIZ} we discuss events outside our flare sample, which appear extreme enough to cause large-amplitude changes in the \(iz\) bands.

{\refbf Our ability to discover extremely bright flares is limited by the rejection of saturated measurements from the parent data set. The filters reaching the most shallow magnitudes are $u$ and $v$, with an average saturation limit of 9~mag. Most stars are at the faint end with quiescent magnitudes of $u=16\ldots 18$, where saturation limits us to flares of $\Delta u<7\ldots 9$. This implies that we are largely sensitive to much more extreme flares than the brightest single event in our sample, which has $\Delta u\approx 4.7$ (see Fig.~\ref{fig:flare amplitudes}). Also, the single brightest star in our sample has $u\approx 14$ and $v\approx 14$, which limits the dynamic range for flare detection to $\Delta u<5$. We thus consider it unlikely that we have missed an extreme flare due to saturation.}

\subsection{Flare spectral energy distribution from SkyMapper {\it uvgriz} photometry}
Spectral energy distributions (SEDs) of flares provide important constraints for their physical models, their simulation and radiation mechanisms. In the past, M dwarf flares have {\refbf mostly} been observed in {\refbf Johnson} \(U\) band (3300--4000\AA ) where they show the highest contrast {\refbf within} the spectral range {\refbf of CCDs}. {\refbf Some observations of flares cover} the {\refbf full range of} \(UBVR\) {\refbf bands} and suggests that flares around peak {\refbf resemble} a featureless, blackbody-like emission with temperatures of $T\sim$9\,000--10\,000 K \cite[e.g.][]{Hawley1992,Hawley2003,kow2013}. However, it is difficult to disentangle the relative contributions from higher-order Balmer lines in broadband filters \citep{Allred2006,Kowalski2019}. Here, the SkyMapper filters {\refbf might have an} advantage as they split the \(U\) band into two, the ultraviolet \(u\)~band (centre: 349nm/width: 42nm) and the violet \(v\)~band (384nm/28nm). {\refbf In the following, we will investigate this possibility.}

\begin{figure}
\begin{center}
\includegraphics[width=\linewidth]{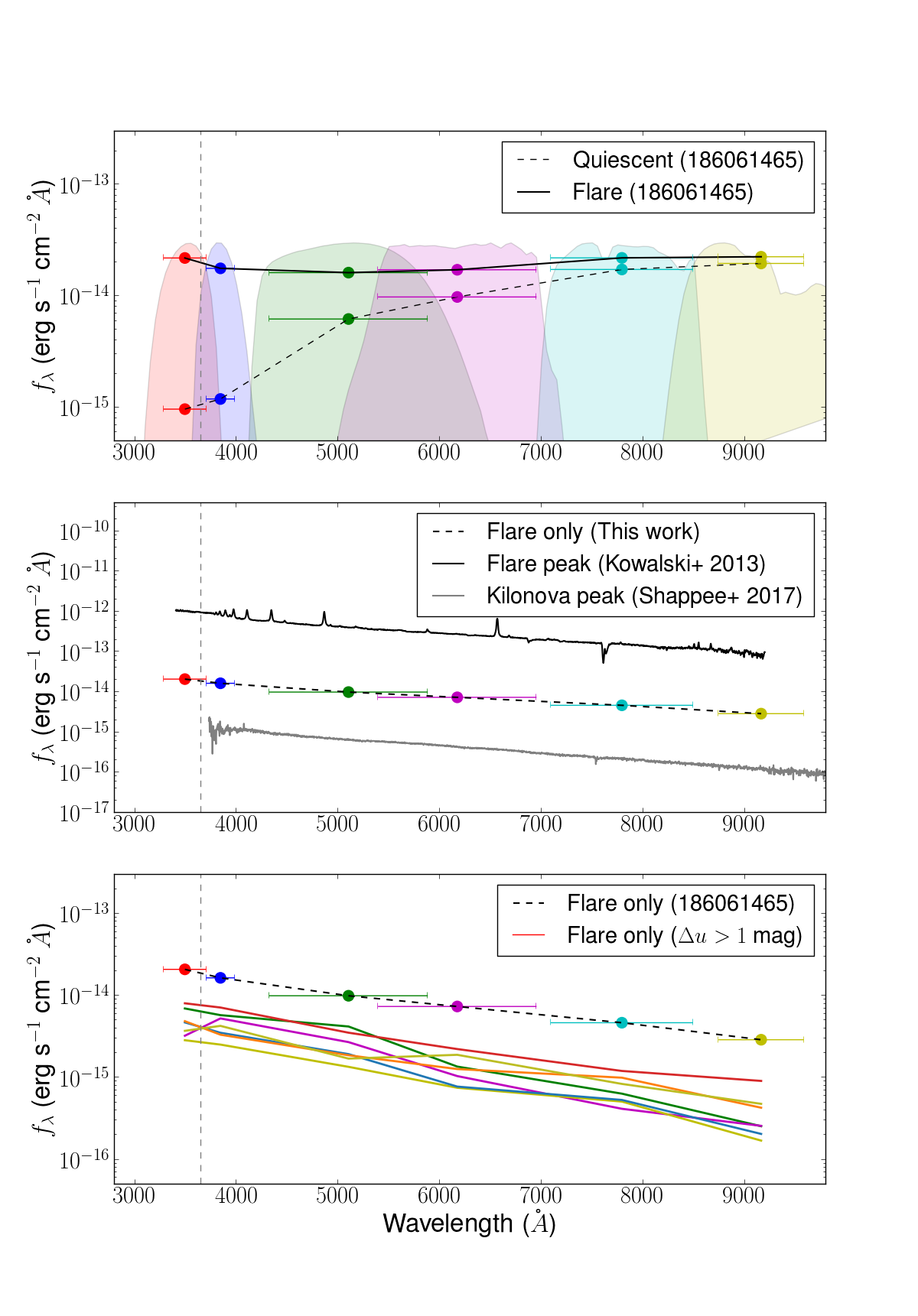}
\caption{Top: spectral energy distribution of an M dwarf (DR1 \texttt{object\textunderscore id} = 186061465; \texttt{SMSS J134839.68-452710.2}) during a large flare with $\Delta u = 3.4$ (solid line) and in quiescence (dashed line). The vertical line marks the Balmer jump at 3646\AA. {\refbf Normalised transmission curves for SkyMapper \(uvgriz\) filter are shown as lighter shaded regions.} Middle: The flare SED after subtracting the quiescent flux, compared with the spectrum of YZ CMi at flare peak (black line; \citealt{kow2013}) and the 0.49-day spectrum of the kilonova from GW170817 (grey line; \citealt{Shappee2017}). Bottom:  The flare-only SEDs for M dwarf flares with large amplitude ($\Delta u > 1.0$ mag) in our sample.}
\label{fig:Flare SED}
\end{center}
\end{figure}

To obtain flux densities $F_{\lambda}$ in each of the SkyMapper filters during the quiescent and flare states, we convert SkyMapper AB magnitudes into flux densities $f_\lambda$ following the definition of \citet{Bessell2012}. Fig.~\ref{fig:Flare SED} compares the SED of one M dwarf in the non-flaring and flaring state. The former shows a normal M dwarf SED with $T_\mathrm{eff}$ $\simeq$ 3500 K, while the blue upturn in the latter is a clear sign of a composite spectrum. After subtracting the SED in the quiescent state, we compare the {\refbf flare-only} SED with {\refbf a peak-phase} spectrum of a flare on the M dwarf YZ CMi \citep{kow2013}, and with the earliest spectrum of a kilonova observed $\sim$12 hours after a binary neutron-star merger \citep{Shappee2017}. All three appear to be characterised by black-body emission with similar temperatures of $\sim$10,000 K. In the bottom panel of Fig.~\ref{fig:Flare SED}, we show the flare-only SEDs of flares with large amplitude variations of $\Delta u > 1$ mag. These large amplitude flares are likely observed near the peak epoch because all the flare-only SEDs are dominated by hot black-body emission with a temperatures of $\sim$10\,000 K. It is also trivially expected that flares seen near peak will show up preferentially with larger amplitudes.

{\refbf To investigate the flare temperature and mix of radiation mechanisms in detail, we check colour-colour diagrams of flares with large amplitudes ($\Delta u \geq 1$ mag) that could be close to flare peak. Our measurements of these events are all blue and centred on \((u-v)_{0}\approx 0\) and \((v-g)_{0} \approx 0\), indicative of black-body temperatures between 10\,000 K and 20\,000 K. But there is significant scatter around this mean colour, which is caused in part by the short time difference between observations of the $uvg$ filters, during which the flare evolves in brightness. More importantly, it might be caused by the Balmer continuum with a wide range of relative contributions to the continuum flux in the \(uv\) filters. Thus, we cannot determine the exact flare temperature only by colour information without simultaneous spectra.}

\subsection{Flaring fraction of M dwarfs by type and position}
The nearly complete coverage of a whole hemisphere of sky allows us to investigate with high statistical significance the flaring fraction of nearby M dwarfs as a function of {\refbf spectral subtype and position in the Galaxy}. The observed flare volume is unbiased within the {\refbf considered} area (2\,965 SkyMapper fields with $\sim$16\,900 deg$^2$), which covers a wide range in Galactic longitudes ($0\degr$ to $360\degr$) and latitudes ($-90\degr$ to $-10\degr$ and $+10\degr$ to $+70\degr$). We avoid higher-reddening regions as these are coincident with crowded regions, where it is difficult to measure accurate photometry. In this section, we will derive robust statistics of flare rates by grouping stars by their properties with at least 1\,000 stars per bin and measuring the flare fraction from our flare sample.

\begin{figure}
\begin{center}
\includegraphics[width=\linewidth]{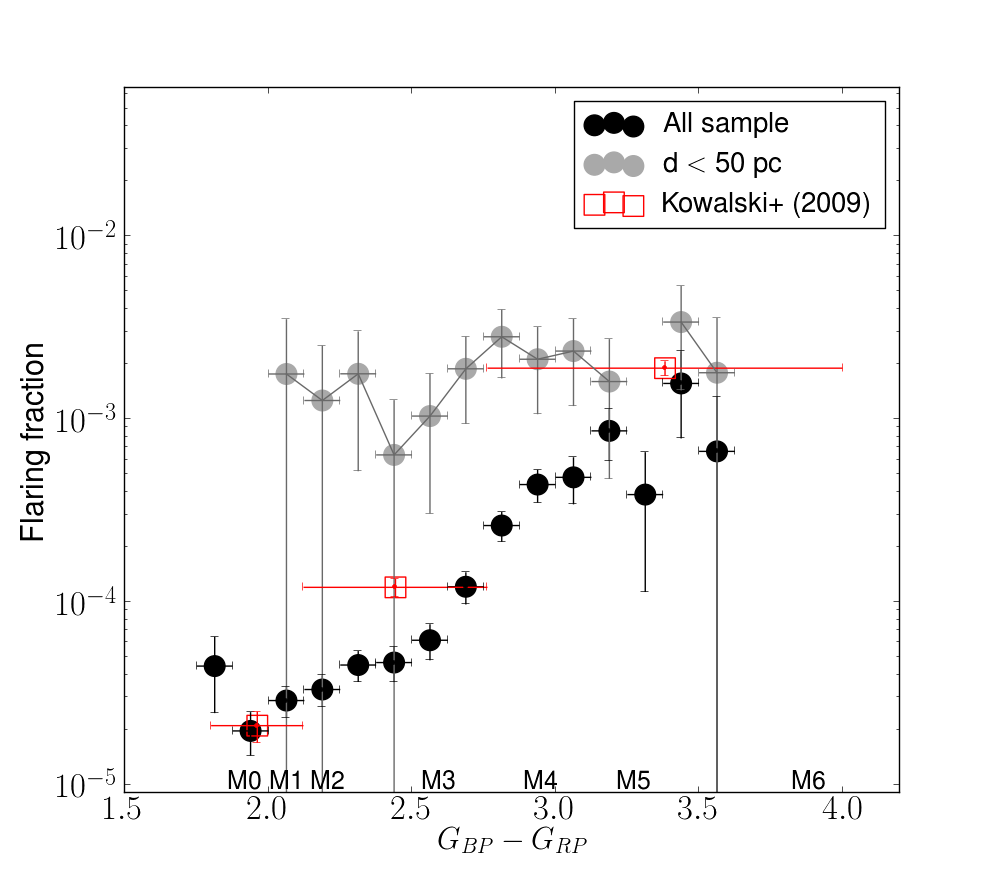}
\caption{Flaring fraction vs. colour: horizontal error bars are bin width and vertical errors bars are Poisson uncertainties. Spectral types mark the colours of \citet{Pickles1998} dwarf templates. The \citet{kow2009} points {\refbf (open squares)} have been placed to match the spectral types {\refbf and are consistent with our full flare sample (black symbols). Grey symbols show a volume-limited sample within 50~pc distance from the Sun.}} \label{fig:Flaring fraction vs. Gaia colour}
\end{center}
\end{figure}

We investigate the fraction of M dwarfs that flare as a function of Gaia colour ($G_\mathrm{BP}-G_\mathrm{RP}$) in bins of 0.125~mag. The flare fraction is defined as the ratio between the number of flare epochs and the total number of observed M dwarf epochs (flaring or not), or in other words, the fraction of stars seen as flaring in an instantaneous snapshot of the sky. This is shown in Fig.~\ref{fig:Flaring fraction vs. Gaia colour}, where we see a clear trend of increasing flaring fraction towards redder colour {\refbf in the black points that represent the whole sample}. The flaring fraction rises from $\sim 30$ flares per million stars for early M dwarfs ($G_\mathrm{BP}-G_\mathrm{RP}$ $\sim 2$) to $\sim $1\,000 flares per million stars for late M dwarfs ($G_\mathrm{BP}-G_\mathrm{RP}$ $\sim 3.5$). We place sub-type labels in Fig.~\ref{fig:Flaring fraction vs. Gaia colour} based on synthetic photometry of the \citet{Pickles1998} templates. 

We compare this result to a flare sample of nearly equal size \citep{kow2009} that was obtained in the much smaller-area SDSS Stripe 82 from repeat observations over roughly 60 epochs. This latter sample covers about 280 deg$^{2}$ along the celestial equator stretching in Galactic longitude across $45\degr <l< 191\degr$ and in latitude across $-64\degr <b< -21\degr$. Our distribution is similar to {\refbf theirs}, but provides tighter constraints for the flaring fraction as a function of colour. 
\begin{figure}
\begin{center}
\includegraphics[width=\linewidth]{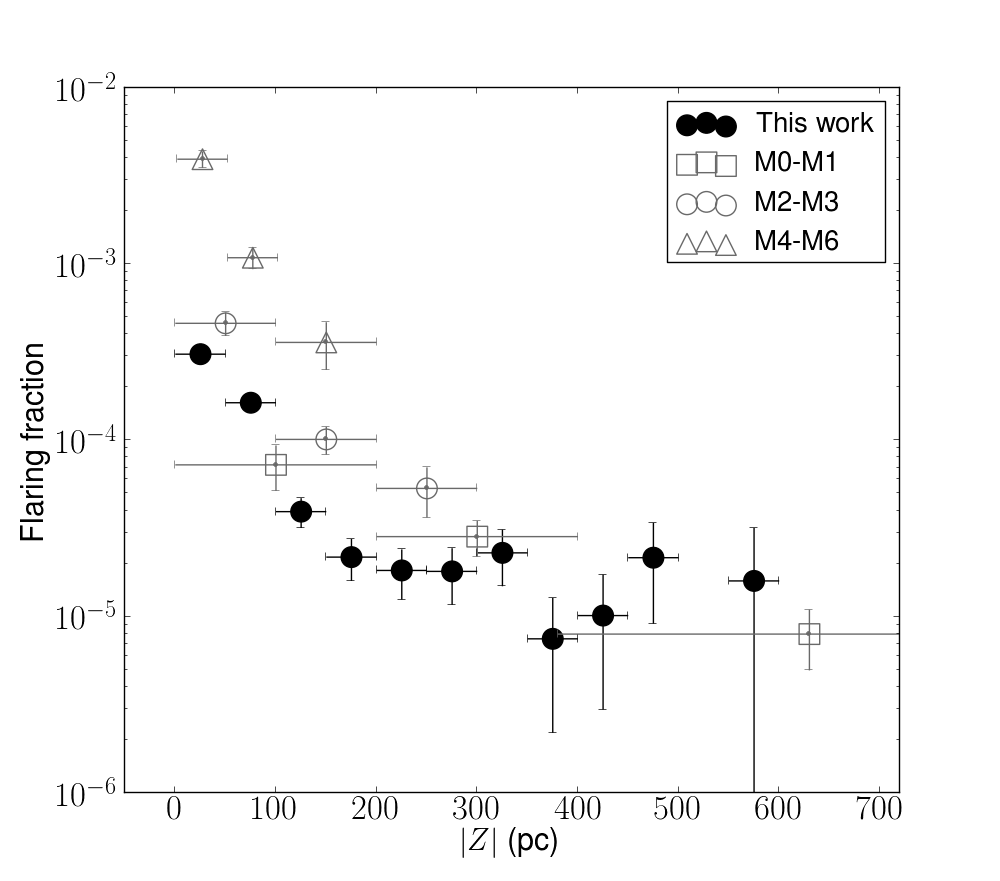}
\caption{Fraction of flaring epochs as a function of vertical distance, $|Z|$, from the galactic plane (black points, {\refbf $|Z|$ bin size 50 pc). Error bars are Poisson uncertainties in the flaring fraction and bin width in $|Z|$. We overlay results from \citet{kow2009} for their spectral type ranges (M0--M1: open squares, M2--M3: open circles and M4--M6: open triangles)}.}
\label{fig:Duty cycle vertical distance}
\end{center}
\end{figure}

\begin{figure*}
\includegraphics[width=0.33\linewidth]{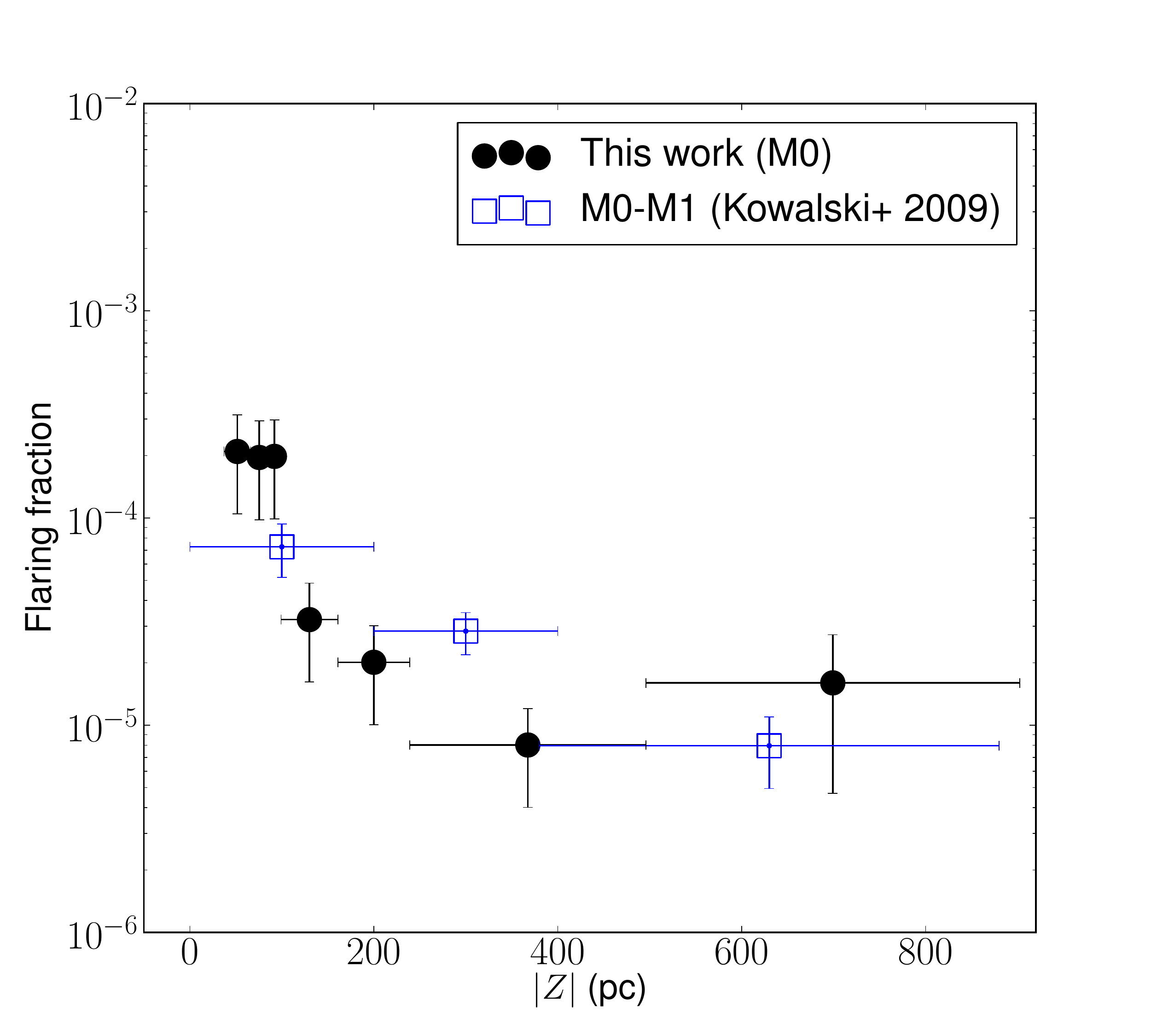}
\includegraphics[width=0.33\linewidth]{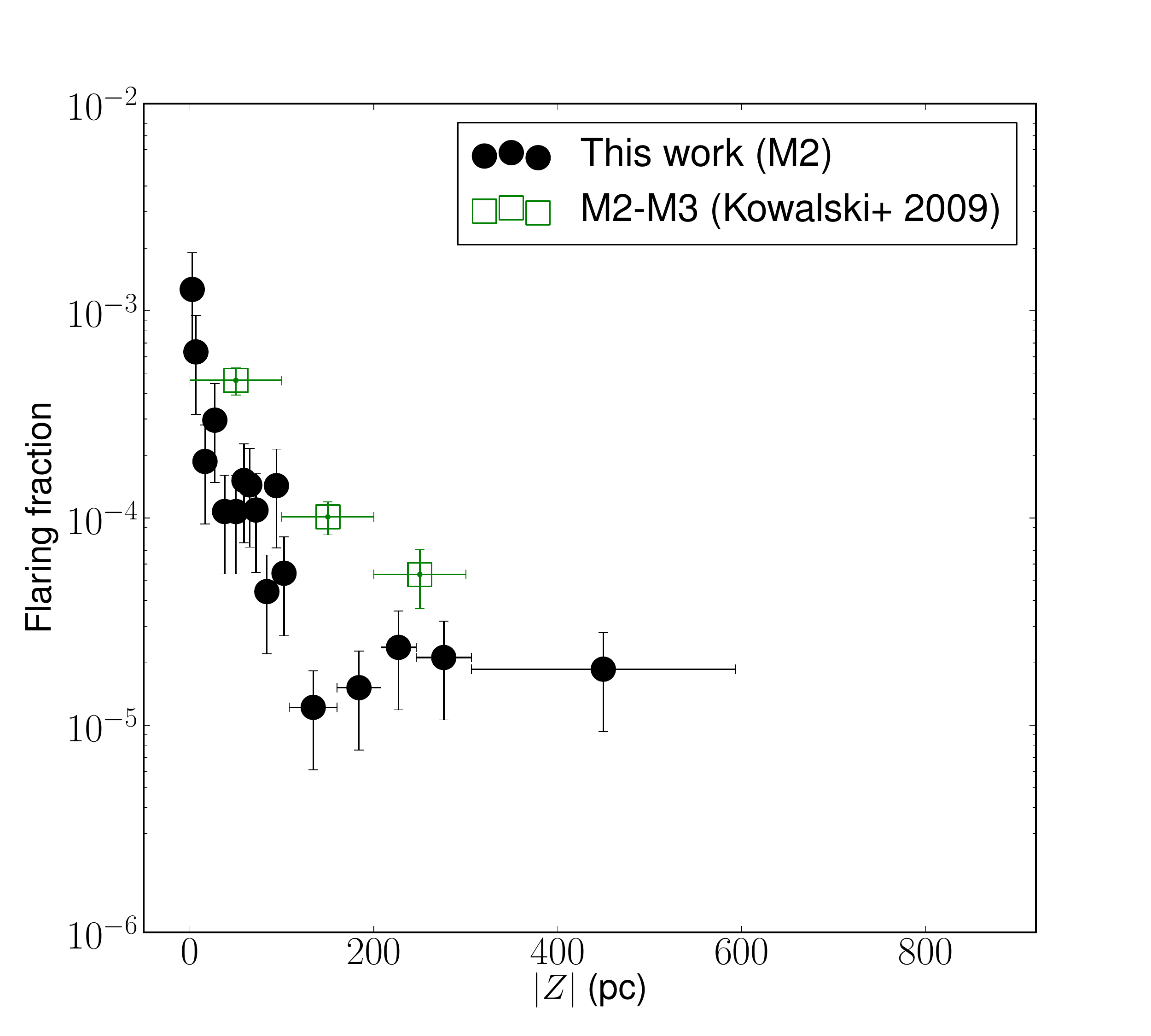}
\includegraphics[width=0.33\linewidth]{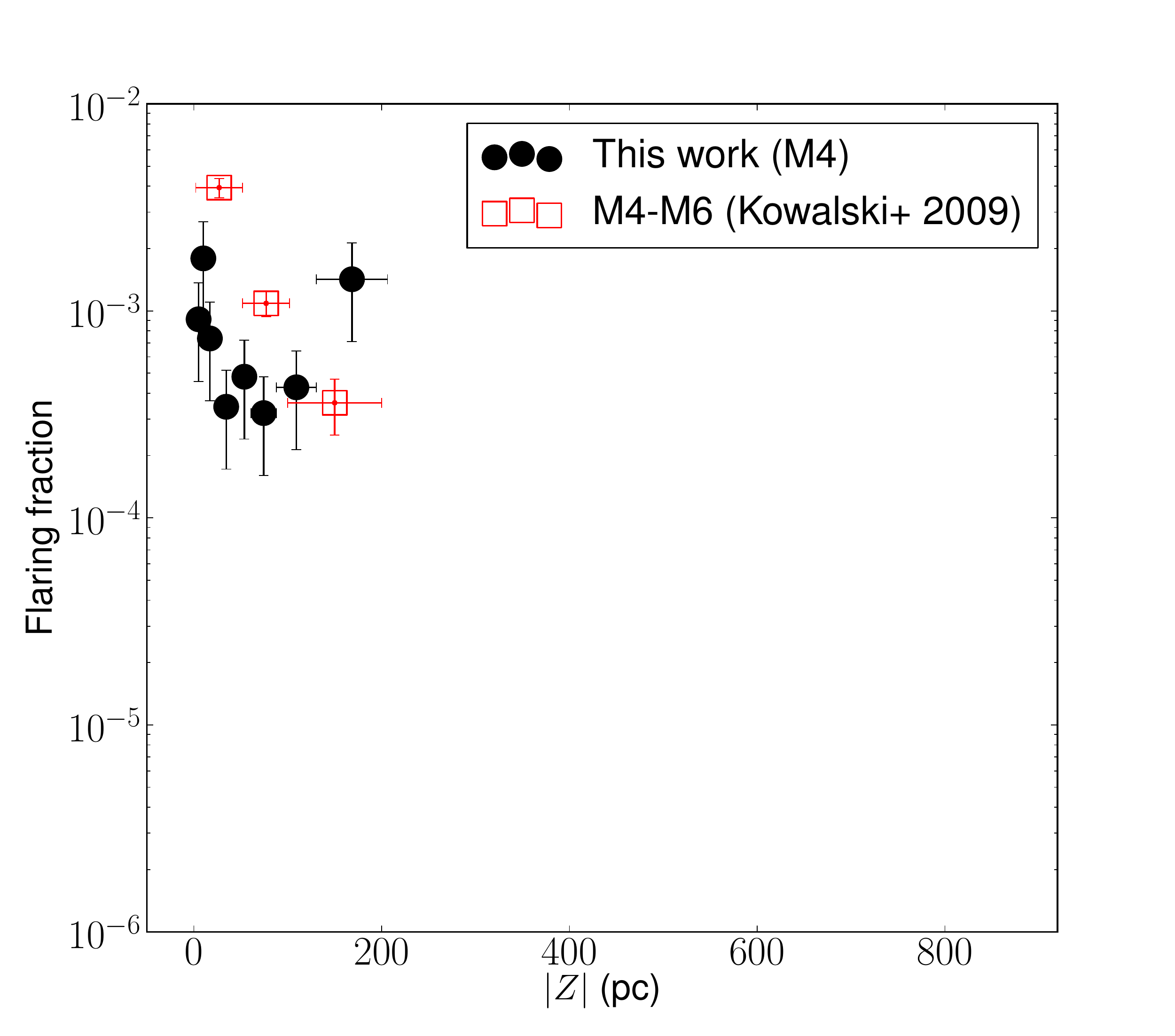}
\includegraphics[width=0.33\linewidth]{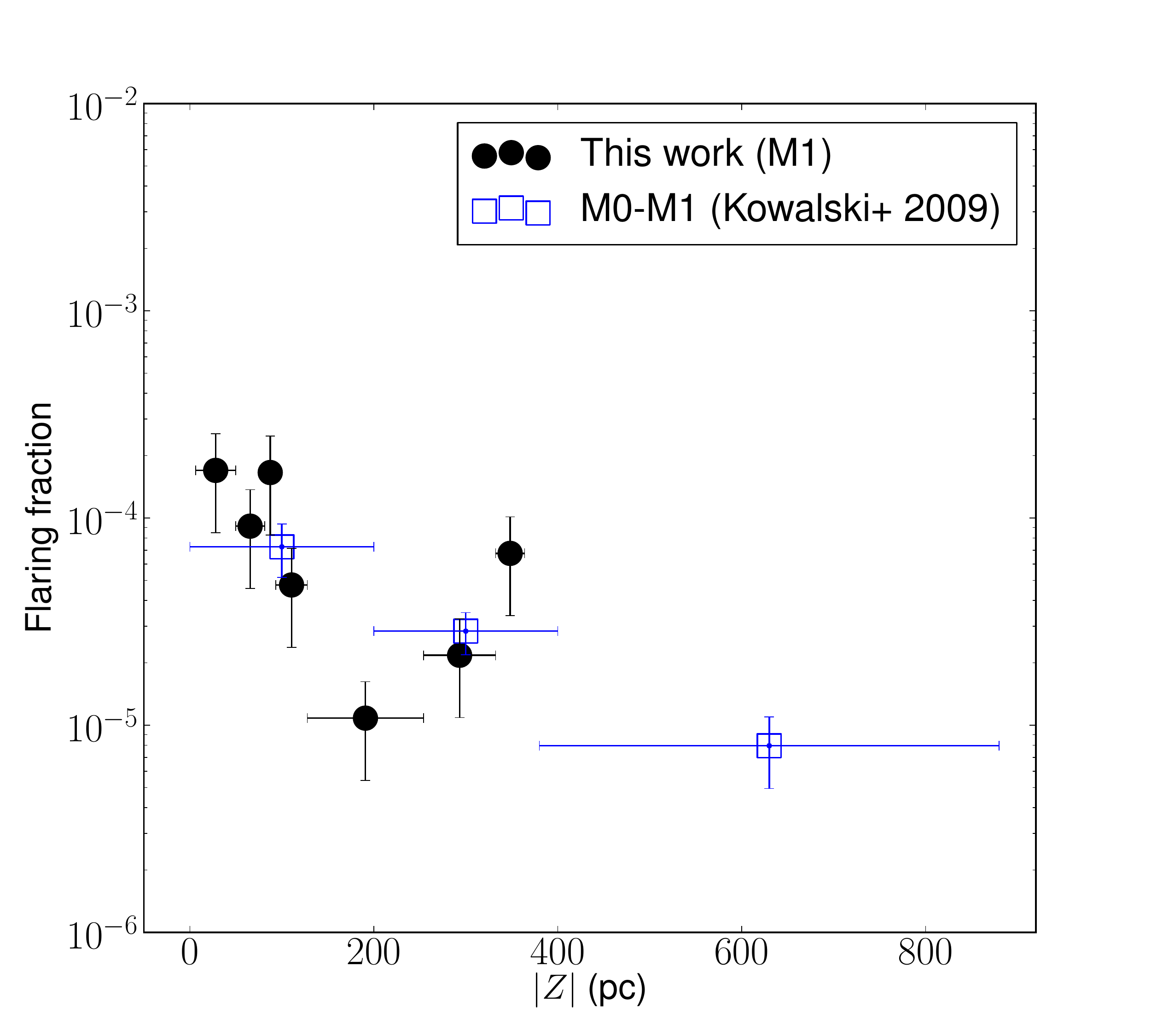}
\includegraphics[width=0.33\linewidth]{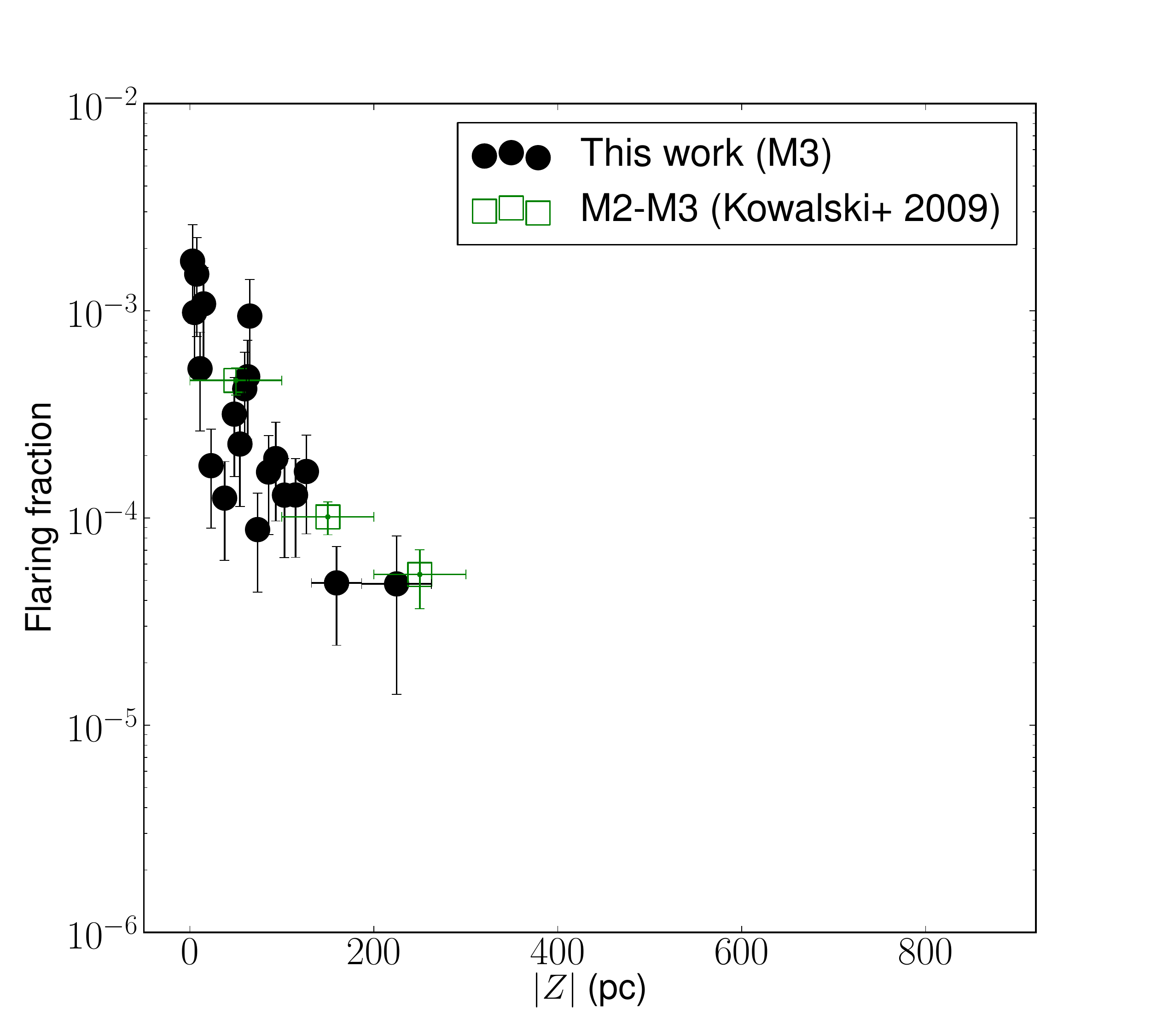}
\includegraphics[width=0.33\linewidth]{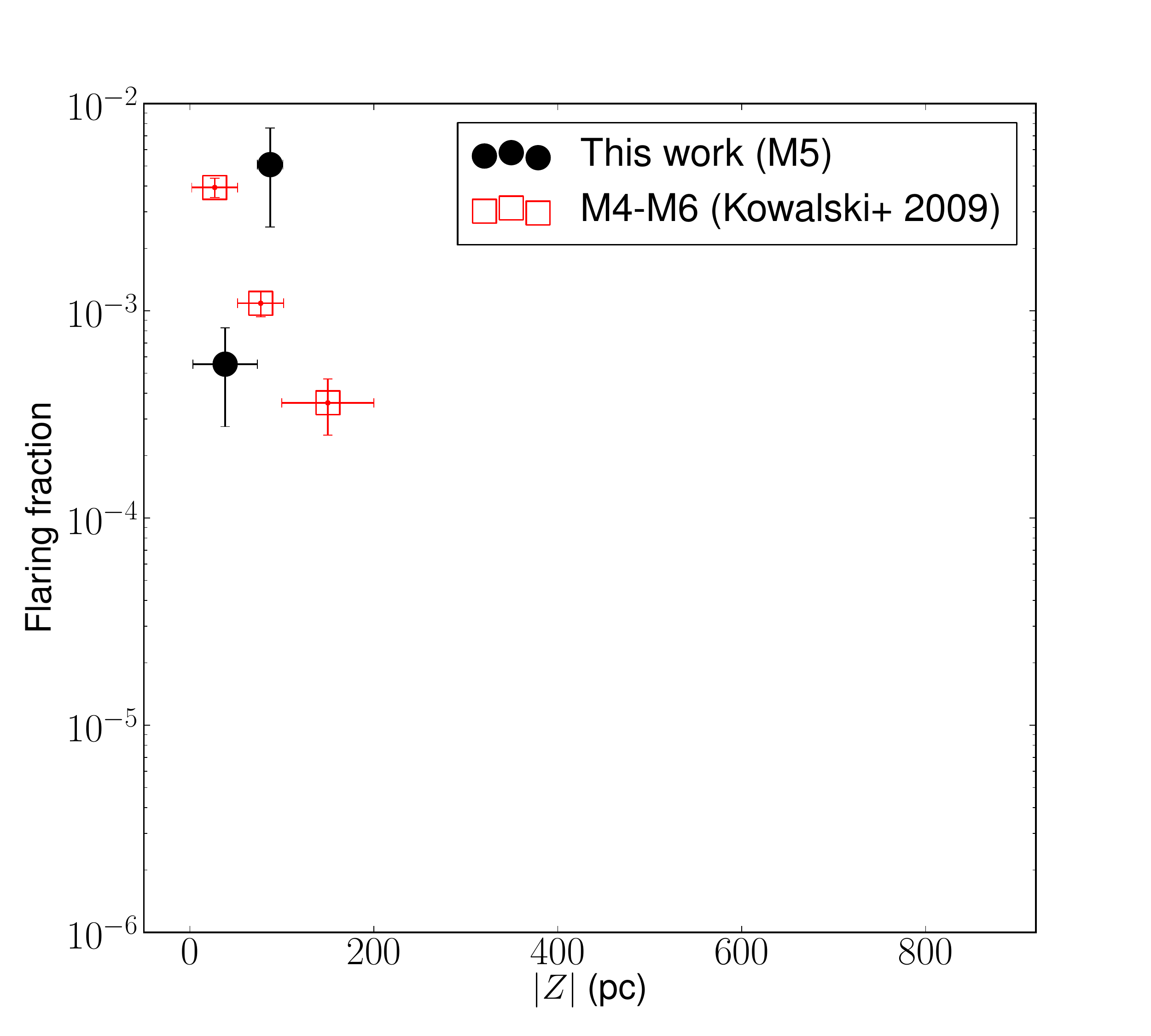}
\caption{Same as Fig.~\ref{fig:Duty cycle vertical distance} but showing the results for spectral subtypes from M0 (top left) to M5 (bottom right). {\refbf The width of the $|Z|$ bins is adjusted so that they include a constant number of flare events.} }
\label{fig:Flaring fraction vs. spectral types}
\end{figure*}

One interesting feature in our distribution is that the flaring fraction {\refbf appears to increase} steeply between M3 and M4 spectral type. {\refbf Perhaps by coincidence, this is also} the regime where the interior structure of main-sequence dwarf stars is thought to transition from partially to fully convective and where an abrupt change in the large-scale magnetic topology is expected (see \citealt{Stassun2011} and references therein). {\refbf Also}, \citet{Rabus2019} identified a sharp transition region in the $T_\mathrm{eff}$--radius plane between partially and fully convective M-dwarfs, {\refbf which} appears at a magnitude $M_\mathrm{G} \sim$11--11.5 and {\refbf a colour of} $G_\mathrm{BP} - G_\mathrm{RP}$ =  2.55--3.0 (see fig.~9 of \citealt{Rabus2019}). {\refbf For these reasons, the feature in the trend may appear to be physically meaningful.}

{\refbf However, this is also the colour range where our search volume changes from including the thick disc and halo to only the thin disc, so that the age distribution of our stars changes rapidly across the M3-to-M4 range. For this reason, we also consider the flare fraction in a smaller sample limited to a small vertical distance from the plane of the Galaxy of $|Z|<50$~pc, where we expect little age gradient within the volume. But in that sample we find at low Galactic latitude that we reach large distances for the stars, but only closer distances for the detected flares, so that the sensitivity of flare detection is still correlated with spectral type. 

Then we consider a volume-limited sample of stars and flares within a solar neighbourhood of $d<50$~pc, which is not only expected to be free of age gradients but also statistically more complete in its flare detection. This sample includes 5\,889 stars with 18\,979 observation blocks and 32 flares. The results of the volume-limited sample are shown as grey points in Fig.~\ref{fig:Flaring fraction vs. Gaia colour} and surprisingly show no dependence on spectral subtype -- instead we find a constant flaring fraction of $\sim $1\,650 per million M~stars across the types from M1 to M5. Given that our flare detection is mostly limited by changes in $u$ band magnitude, this finding suggests that flare rates are constant when selecting flares by power relative to $u$ band luminosity of the star, i.e. we find qualitatively that ``strong flares on bright stars are as common as weak flares on faint stars''. Going towards warmer stars, we still expect the flare rate to drop eventually, but we don't see the turnover in the range of M dwarfs.}

We also investigate the flaring fraction {\refbf as a function of the} vertical distance $|Z|$ from the Galactic plane {\refbf and thus as a function of stellar age}. Following \citet{Gaia2018A&A...616A..11G}, we compute the Galactic cylindrical coordinates ($R,\phi,Z$) where the radial distance of the Sun from the Galactic centre is assumed as $R_{\odot}=8.34$~kpc and the height of the Sun above the plane is $Z_{\odot}=27$~pc. As in the SDSS flare sample, the range of vertical distance in the sample varies with the observed brightness of the stars. In our sample, early M dwarfs ($G_\mathrm{BP} - G_\mathrm{RP}$ = 2) are visible up to $|Z|=1.5$~kpc, while late M dwarfs ($G_\mathrm{BP}-G_\mathrm{RP}$ = 3.5) are confined to within $|Z|=200$~pc. Fig.~\ref{fig:Duty cycle vertical distance} shows a strong dependence of the flaring fraction on $|Z|$, similar to the trends seen in earlier photometric and spectroscopic samples (e.g., \citealt{kow2009, hilton2010, west2011}). Since the slope of this relation varies as a function of spectral type, we show {\refbf the SDSS} results for three different spectral ranges. Note that distances to the SDSS stars were determined with a photometric parallax relation and are thus less certain and not necessarily consistent with our distance scale derived from Gaia. The mean flaring fraction in our combined sample is consistent with early M dwarfs (M0-M1) in Stripe 82, and these are the dominant objects in our flux-limited sample. We also find that M dwarfs within 150 pc of the Galactic plane show relatively larger flaring fraction than those farther from the plane, which seem to converge towards a level of 10 to 20 per million. This decline is usually explained by a rapidly declining magnetic activity in older stars \citep{west2006,West2008}, which were born in the Galactic disc a long time ago, but were dynamically heated by gravitational interaction and scattered into orbits with larger mean $|Z|$. 

Lastly, we check the results on the $|Z|$ dependence by spectral subtype to see if the trend is {\refbf an} artefact of mixing spectral types. {\refbf In Fig.~\ref{fig:Flaring fraction vs. spectral types} we show the dependence on $|Z|$ for six subtypes in our sample, from M0 to M5. We choose the bin width in $|Z|$ such that we have} the same number of flares in each bin. {\refbf Where we have sensitivity, we see that the flaring fraction closest to the plane of the Milky Way is over 1\,000 per million stars, while at more than 150~pc vertical distance it has dropped to a few tens per million stars and tends to stay flat beyond. These trends are fairly similar across the subtypes from M0 to M3 at least.}

\begin{figure*}
\includegraphics[clip=true,width=0.49\linewidth]{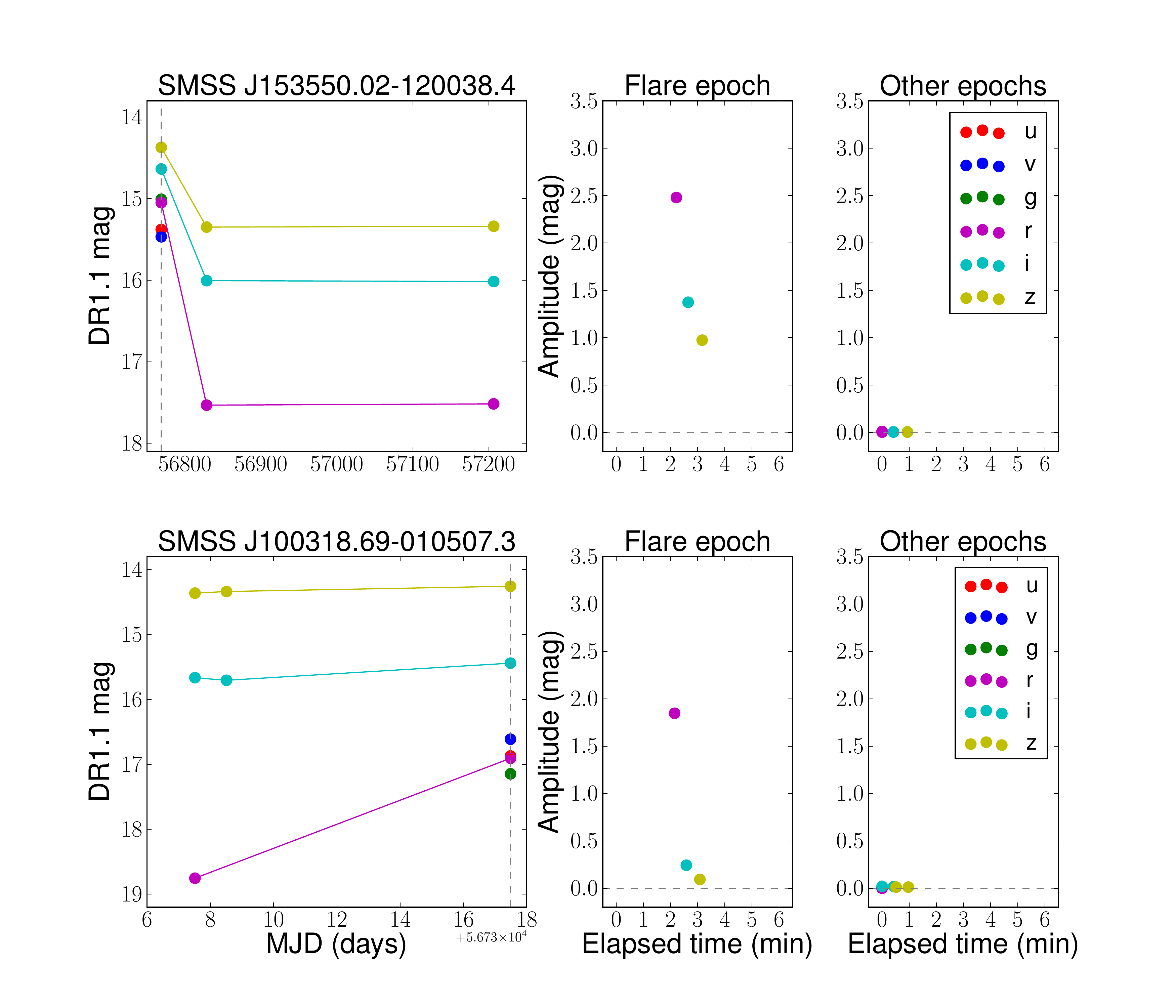}
\includegraphics[clip=true,width=0.49\linewidth]{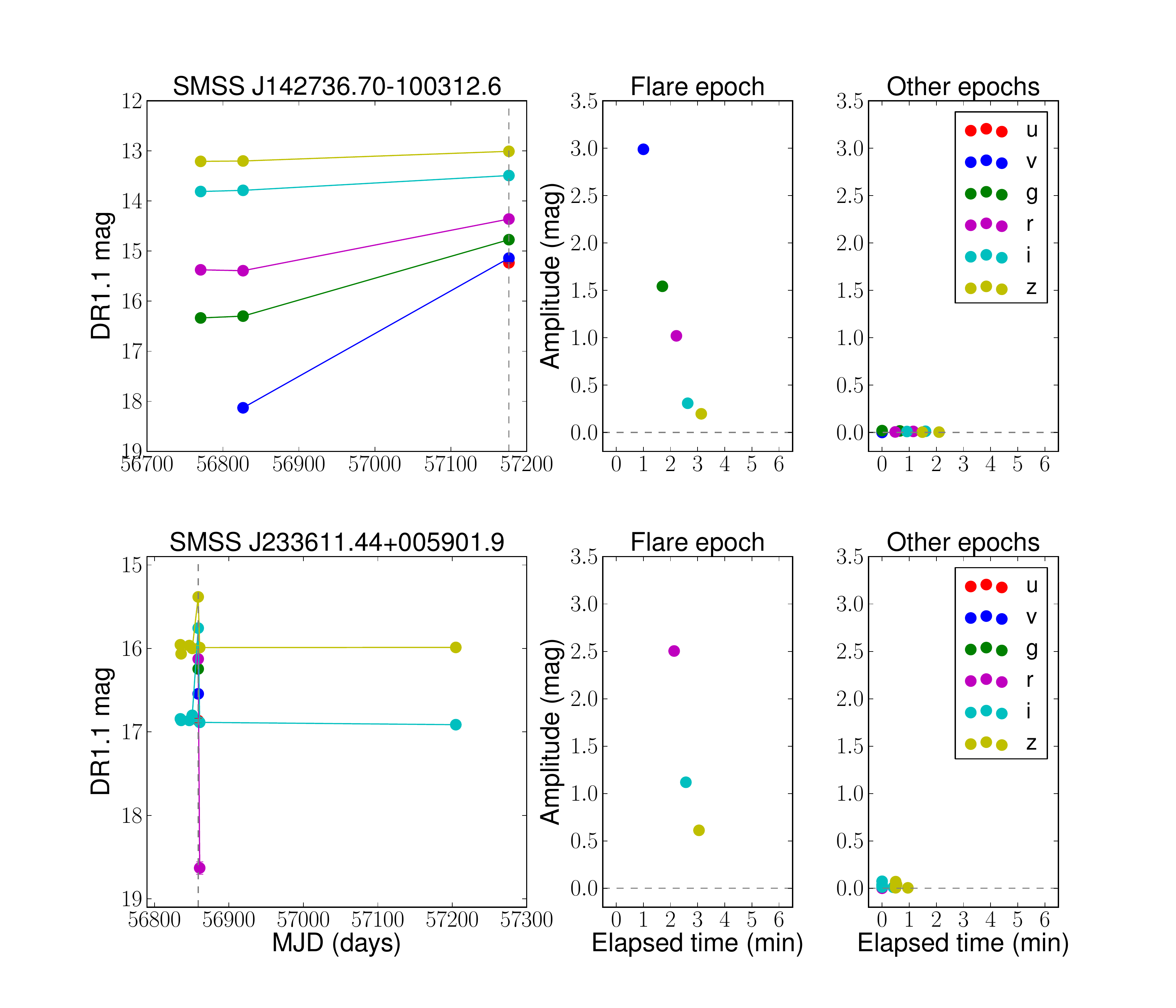}
\caption{Light curves of four high-amplitude flares with large continuum enhancement in the redder filters. The panels to the right show a zoom-in of the flaring epoch and a non-flaring epoch for comparison. The objects are labelled with their SMSS object IDs.
}\label{fig:Lageamplitude flares}
\end{figure*}

\begin{table*}
\caption{Flaring fraction for M dwarfs observed in different ranges of Galactic latitude.}
\centering
\begin{tabular}{ccccc}
\hline \noalign{\smallskip}
Galactic latitude & Number of stars & All observed blocks & Flaring blocks &  Flaring fraction $f$ per million stars \\
\noalign{\smallskip} \hline \noalign{\smallskip}
$10\degr\leq |b| <15\degr$  &  90,996 &  261,651 &  24  &  $91.7\pm18.7$ \\ 
$15\degr\leq |b| <30\degr$  & 423,294 & 1,281,113 &  78  &  $60.9\pm6.9$   \\ 
$30\degr\leq |b| <45\degr$  & 385,349 & 1,217,773 &  66  &  $54.2\pm6.7$   \\ 
$45\degr\leq |b| <60\degr$  & 281,214 &  936,723 &  50  &  $53.4\pm7.5$   \\ 
$60\degr\leq |b| <75\degr$  & 162,632 &  536,516 &  31  &  $57.8\pm10.4$   \\ 
$75\degr\leq |b| <90\degr$  & 42,629 &  128,038 &   5  &  $39.1\pm17.5$   \\ 
\noalign{\smallskip} \hline \noalign{\smallskip}
Total ($15\degr\leq |b| <90\degr$)  & 1,295,118 & 4,100,163 & 230  &  $56.1\pm3.7$  \\ 
Total ($10\degr\leq |b| <90\degr$)  & 1,386,114 & 4,361,814 & 254  &  $58.2\pm3.6$  \\ 
\noalign{\smallskip} \hline
\end{tabular}
\label{tab:tab3}
\end{table*}

\subsection{Extreme flare candidates in red passbands}\label{LargeFlaresIZ}
We investigate extreme flare cases in the \texttt{master} table, where the quiescent star is not visible in the \(uvg\) bands. We define this subsample as one where we detect only the flare epoch in the bluer filters (\texttt{\{F\}\textunderscore ngood=1}), while the \(r\)~band detections are marginal (\texttt{r\textunderscore ngood=2}) and the \(iz\) detections are regular (\texttt{\{F\}\textunderscore ngood>2}) to get reliable magnitudes in quiescence for applying at least the colour cut $0.25 < i-z < 1.7$. We find 134 objects that are not included in our main analysis, three of which appeared to show flares according to our selection criteria. These three objects show exceptionally large variations of $\Delta m >1$~mag in the \(i\)~band, but two of them were identified as Mira-type stars later with VSX matches.

{\refbf The only plausible extreme flare is} an \(i\)~band flare with $\Delta i$=1.1 mag from the M dwarf \texttt{SMSS J233611.44+005901.9} with estimated flare magnitudes of $>$5.1, $>$5.1, 3.4, 2.5, 0.6 mag for the \(uvgrz\) bands, respectively. {\refbf Fig.~\ref{fig:Lageamplitude flares} shows this flare alongside three weaker ones}. Now, that DR2 of SkyMapper is available \citep{Onken2019} and includes deeper images {\refbf one third} of the {\refbf Southern} sky, we {\refbf can obtain} more precise colour indices of this object in quiescence, $(r - i, i - z) = (2.295, 0.917)$, which indicate a spectral type of M4--M5 that lies on the locus of typical M dwarf stars ({\refbf see} Fig.~\ref{fig:SMSS Mdwarf selection}). For comparison, following the method outlined in \citet{davenport2012}\footnote{\url{https://github.com/jradavenport/flare-grid}}, we simulate a 1.1 mag \(i\)~band flare on an M4 star {\refbf and thus} predict amplitudes in the SDSS \(ugriz\) filters {\refbf of} (6.9, 3.7, 2.2, 1.1, 0.6 mag), {\refbf to which our observed flare amplitudes are well matched}. {\refbf The shape of the SED is perfectly consistent with an M dwarf flare, making it unlikely that} this transient is {\refbf caused by a blend with} an unknown asteroid. 

In Fig.~\ref{fig:Lageamplitude flares}, we highlight the light curves of rare, high-amplitude flares, which have either \(r\) or \(i\) amplitudes of $\geq$ 1 mag. Only one of them, \texttt{SMSS J100318.69-010507.3}, has a Simbad spectral classification (M7Ve) with a strong H$_{\alpha}$ emission \citep{Gizis2002}, further supporting evidence of a young, active star (see bottom left panel of Fig.~\ref{fig:Lageamplitude flares}). As expected, extreme flares are very rare in the \(iz\) filters, with a flaring fraction of $\sim 0.5$ per million stars.

\section{The foreground fog and searches for rare events}
M-dwarf flares are the dominant source of transients in time-domain surveys. We now explore how they might contaminate searches for rare extragalactic transients. A particular {\refbf scenario} is large-area searches for optical counterparts to gravitational-wave (GW) events {\refbf  detected by} the LIGO/Virgo {\refbf facilities} \citep{Abbott2017,Andreoni2017}, with the aim of identifying potential kilonova emission as early after the GW event as possible. Typical cases may involve searching an area of 100 square degrees.

To estimate the predicted rate of M dwarf flares for the kilonova search by SkyMapper, we first calculate the mean observed fraction of M dwarf flares above our detection limits as a function of Galactic latitude $|b|$. Table \ref{tab:tab3} shows that the flaring fraction $f$ is mostly constant, except for {\refbf low latitudes}, where the mean age and the mean flare rate of thin-disc M dwarfs suddenly changes. The highest-latitude bin appears to show a lower fraction, but {\refbf given} its larger error bar this is insignificant. {\refbf Overall, we find a flaring fraction of $\sim 56$ per million stars at $|b|>15$, while close to the plane this fraction might increase}. 

\begin{figure}
\includegraphics[width=\linewidth]{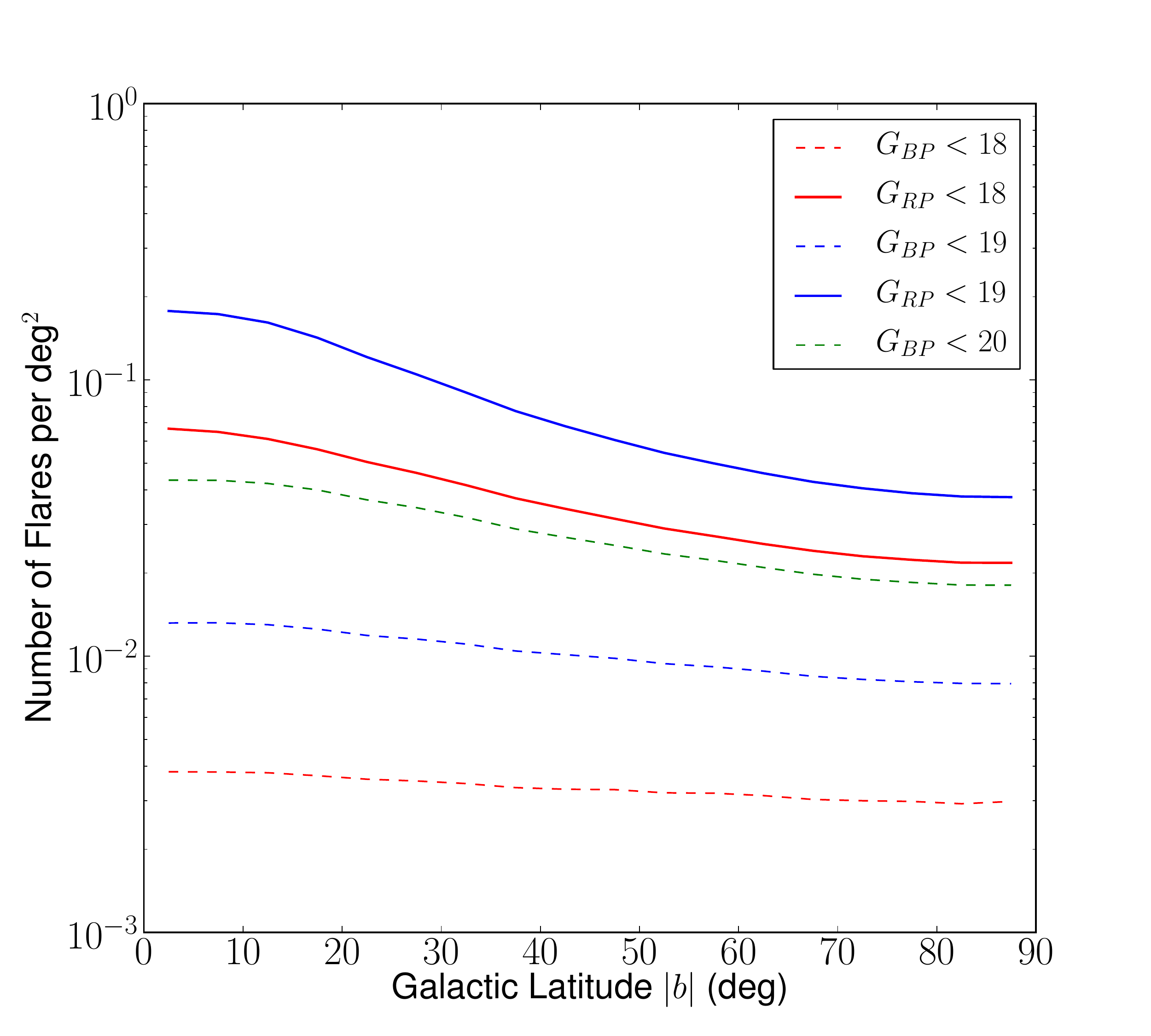}
\caption{{\refbf Surface density} of M dwarf flares with $\Delta u>0.1$~mag for {\refbf samples of different depth} in Gaia  passbands.}\label{fig:Flare fraction with the GUMS-18}
\end{figure}

We now use the most up-to-date simulation of the Gaia Universe Model Snapshot (GUMS-18: \citealt{Robin2012}) to select M dwarfs above a given detection limit and to estimate how many flares a time-domain survey will see instantaneously. This simulation contains about 2.1 billion objects uniformly distributed in the sky that can be observed by Gaia, down to a limiting magnitude of $G=20$. We extract all M dwarfs ranging from spectral type M0V to M9V based on the Gaia Object Generator (GOG18: \citealt{Luri2014}) catalogue, which is the transformation of GUMS-18 into Gaia final catalogue data. We divide the total M dwarf sample into three subsamples covering different magnitude cuts (18, 19 and 20th mag) in each Gaia band. The cumulative number of M dwarfs in the red band, N ($< G_\mathrm{RP}$ = 18--19, which includes 32--76M detectable objects), is an order of magnitude larger than that in the blue band (2--8M objects). {\refbf We drop the case of $G_{RP}<20$ because the} red sample is incomplete at $G_{RP} > 19$. 


According to the Gaia simulation, the number density of $G_{RP}$~band M dwarfs is decreasing with increasing latitude $|b|$, ranging from a few thousand to a few hundred objects per square degrees. We multiply the number density with the flaring fraction derived by this work to obtain an instantaneous rate of flares that would be found in a single visit. Fig.~\ref{fig:Flare fraction with the GUMS-18} shows the expected number of flares from M dwarfs brighter than a given Gaia magnitude limit per square degrees as a function of galactic latitude. Due to the assumed flat slope of our observed flaring fraction with galactic latitude, the overall distribution is primarily driven by number density rather than physical changes. {\refbf Although the flare rate of M dwarfs has a strong dependence on spectral type and age, our results are still strongly dominated by the number density of M dwarfs for each spectral type. In the Gaia simulation, more than 80 and 16 percent of the objects in the sample are early-type (M0-M1) and mid-type M dwarfs (M2-M3), respectively. The number of late-type M dwarfs (M4-M5) is more than two orders of magnitude below the dominant sample.} We can compute the expected number of flares from M dwarfs that would be seen down to a given Gaia magnitude during a {\refbf hypothetical} snapshot observation that covers 100 square degrees (Table \ref{tab:tab4}). Over this area we expect to observe about one flare with $\Delta u>0.1$~mag from M dwarfs with $G_\mathrm{BP} < 19$, whereas we expect 4 to 18 flares from M dwarfs with $G_\mathrm{RP} < 19$, depending on the line of sight through the Galaxy. 

How many flares will be seen by a survey within a given magnitude limit depends on the flare contrast and thus on the observed passband. While there are many more M stars seen in quiescence in redder bands, the relative amplitude of flares is much smaller than in blue passbands. Given the flare temperature, flaring stars can appear brighter in the blue than in the red during the rare extreme flares. Therefore, a prediction of expected flares in observations with blue passbands would need to consider much fainter M dwarfs than considered in the Gaia simulation. This means that searches for transients in the blue will be dominated by flares that appear to come out of nowhere, as blue images will not show the stars they originate from in quiescence. While a detailed interpretation of this contamination as a function of apparent peak magnitude of the flare is beyond the scope of this work, we do note, that knowing about the presence of an M dwarf at the location of a new transient, from redder images or the Gaia catalogue itself, will allow the transient to be identified as an M-dwarf flare. 

Transient searches in red bands, however, will show the M stars in quiescence and render the flare just as an event of moderate variability. For example, SkyMapper's programme to search for kilonovae from GW events uses the $i$~band. As shown in Sect.~\ref{LargeFlaresIZ}, strong flares with $\Delta i>1$~mag are two orders of magnitude less common than the flares with $\Delta u>0.1$~mag underlying the above model prediction. Hence, {\refbf during a realistic kilonova search with SkyMapper we expect no flare to appear from M dwarfs that were not already seen by the Southern Survey} in quiescence. 

\begin{table}
\caption{Expected number of flares from M dwarfs with magnitude brighter than a given Gaia magnitude limit per field of 100 square degrees.}
\centering
\begin{tabular}{ccccc}
\hline \noalign{\smallskip}
& & \multicolumn{3}{c}{Number of flares} \\
\noalign{\smallskip} \cline{3-5} \noalign{\smallskip}
Gaia filter & Mag. limit & $|b|\sim0\degr$ & $|b|\sim45\degr$ & $|b|\sim90\degr$ \\
\noalign{\smallskip} \hline \noalign{\smallskip}
$G_{BP}$ & $< 18$ &  0.4 &   0.3  & 0.3 \\ 
$G_{BP}$ & $< 19$ &  1.3 &   1.0  & 0.8 \\ 
$G_{BP}$ & $< 20$ &  4.3 &   2.5  & 1.8 \\ 
\noalign{\smallskip} \hline \noalign{\smallskip}
$G_{RP}$ & $< 18$ &  6.5 &   3.4  & 2.2 \\ 
$G_{RP}$ & $< 19$ & 17.7 &   6.0  & 3.8 \\ 
\noalign{\smallskip} \hline
\end{tabular}
\label{tab:tab4}
\end{table}

\section{Conclusion}
\label{sec:conclusion}
We study the flaring fraction of M dwarfs in the SkyMapper Southern Survey, taking advantage of the near-hemispheric coverage. We consider the long-cadence multiple visits and label single-epoch outliers as flare candidates. SkyMapper's nearly simultaneous six-filter observations during every visit provide robust evidence of flare variability. The main advantage of our blue-filter observations is that they reveal even low-amplitude flares whereas the detection efficiency drops rapidly towards redder wavelengths. We find the following results:
\begin{enumerate}
    \item We find 254 flare events from 1\,386\,114 M dwarfs in the Southern hemisphere, which is one of the largest samples observed by ground-based surveys.

    \item Our single-epoch flare observations show blue colours {\refbf of} \((u-v)_{0}\approx 0\) and \((v-g)_{0} \approx 0\) {\refbf at near-peak luminosities}, which are indicative of black-body temperatures {\refbf not too different from 10\,000 K. However, we can neither constrain exact flare temperatures nor the contribution of different radiation mechanisms from our colour information alone, despite the non-standard filter set of SkyMapper.}
    
    \item Compared to previous work on the dependency of flare activity on spectral type, we present a more detailed picture: 
    {\refbf in our full sample we observe the known} increase in flaring fraction from $\sim 30$ to $\sim $1\,000 flares per million stars, going from type M0 to M5. {\refbf Crucially, we find an unexpected result when considering a volume-limited sample within 50~pc distance from the Sun: there we find essentially no change in the flare fraction from type M1 to M5. Our data is instead consistent with a constant flare fraction of $\sim $1\,650 flares per million M~dwarfs.} 
    
    \item We confirm a strong dependence of the flaring fraction on vertical distance from the Galactic plane, $|Z|$, which has been reported previously. But now that Gaia DR2 provides much improved distance measurements, we can recognise trends with $|Z|$ more clearly than before. Within 150 pc of the Galactic plane, the flaring fraction declines to a few tens per million stars, largely independent of spectral type. Above {\refbf that} the flaring fraction appears to flatten.

    \item Large-amplitude flares with $\Delta i > 1$ mag are very rare with a flaring fraction of $\sim 0.5$ per million stars. Hence, we typically expect no M-dwarf flare to confuse SkyMapper's $i$ band search for kilonovae from GW events.

\end{enumerate}

We will further investigate flares in future data releases of SkyMapper. The deeper Main Survey will let us search a larger volume with more distant M dwarfs. {\refbf It} also includes observing sequences in the $uv$ filters that {\refbf include} three \(u\)--\(v\) pairs taken within a 20 minute-visit to each field, which will show flares evolve in brightness and colour. Future data releases will also improve coverage close to the Galactic plane, enabling investigation of the flare properties of M dwarfs with $|b| < 10\degr$.

\section*{Acknowledgements}
{\refbf We are grateful to an anonymous referee for their comments and ideas that significantly improved the manuscript.}
The national facility capability for SkyMapper has been funded through ARC LIEF grant LE130100104 from the Australian Research Council, awarded to the University of Sydney, the Australian National University, Swinburne University of Technology, the University of Queensland, the University of Western Australia, the University of Melbourne, Curtin University of Technology, Monash University and the Australian Astronomical Observatory. SkyMapper is owned and operated by The Australian National University's Research School of Astronomy and Astrophysics. The survey data were processed and provided by the SkyMapper Team at ANU. The SkyMapper node of the All-Sky Virtual Observatory (ASVO) is hosted at the National Computational Infrastructure (NCI). Development and support the SkyMapper node of the ASVO has been funded in part by Astronomy Australia Limited (AAL) and the Australian Government through the Commonwealth's Education Investment Fund (EIF) and National Collaborative Research Infrastructure Strategy (NCRIS), particularly the National eResearch Collaboration Tools and Resources (NeCTAR) and the Australian National Data Service Projects (ANDS). Parts of this project were conducted by the Australian Research Council Centre of Excellence for All-sky Astrophysics (CAASTRO), through project number CE110001020, and parts were funded by the Australian Research Council Centre of Excellence for Gravitational Wave Discovery (OzGrav), CE170100004.
This work has made use of data from the European Space Agency (ESA) mission {\it Gaia} (\url{https://www.cosmos.esa.int/gaia}), processed by the {\it Gaia} Data Processing and Analysis Consortium (DPAC, \url{https://www.cosmos.esa.int/web/gaia/dpac/consortium}). Funding for the DPAC has been provided by national institutions, in particular the institutions participating in the {\it Gaia} Multilateral Agreement.


\bibliographystyle{mnras}
\bibliography{SMSS_Mdwarf_Flares}

\begin{thebibliography}{}
\makeatletter
\relax
\def\mn@urlcharsother{\let\do\@makeother \do\$\do\&\do\#\do\^\do\_\do\%\do\~}
\def\mn@doi{\begingroup\mn@urlcharsother \@ifnextchar [ {\mn@doi@}
  {\mn@doi@[]}}
\def\mn@doi@[#1]#2{\def\@tempa{#1}\ifx\@tempa\@empty \href
  {http://dx.doi.org/#2} {doi:#2}\else \href {http://dx.doi.org/#2} {#1}\fi
  \endgroup}
\def\mn@eprint#1#2{\mn@eprint@#1:#2::\@nil}
\def\mn@eprint@arXiv#1{\href {http://arxiv.org/abs/#1} {{\tt arXiv:#1}}}
\def\mn@eprint@dblp#1{\href {http://dblp.uni-trier.de/rec/bibtex/#1.xml}
  {dblp:#1}}
\def\mn@eprint@#1:#2:#3:#4\@nil{\def\@tempa {#1}\def\@tempb {#2}\def\@tempc
  {#3}\ifx \@tempc \@empty \let \@tempc \@tempb \let \@tempb \@tempa \fi \ifx
  \@tempb \@empty \def\@tempb {arXiv}\fi \@ifundefined
  {mn@eprint@\@tempb}{\@tempb:\@tempc}{\expandafter \expandafter \csname
  mn@eprint@\@tempb\endcsname \expandafter{\@tempc}}}

\bibitem[\protect\citeauthoryear{{Abbott} et~al.,}{{Abbott}
  et~al.}{2017}]{Abbott2017}
{Abbott} B.~P.,  et~al., 2017, \mn@doi [\apjl] {10.3847/2041-8213/aa91c9},
  \href {https://ui.adsabs.harvard.edu/abs/2017ApJ...848L..12A} {848, L12}

\bibitem[\protect\citeauthoryear{{Allred}, {Hawley}, {Abbett}  \&
  {Carlsson}}{{Allred} et~al.}{2006}]{Allred2006}
{Allred} J.~C.,  {Hawley} S.~L.,  {Abbett} W.~P.,   {Carlsson} M.,  2006,
  \mn@doi [\apj] {10.1086/503314}, \href
  {https://ui.adsabs.harvard.edu/abs/2006ApJ...644..484A} {644, 484}

\bibitem[\protect\citeauthoryear{{Andreoni} et~al.,}{{Andreoni}
  et~al.}{2017}]{Andreoni2017}
{Andreoni} I.,  et~al., 2017, \mn@doi [\pasa] {10.1017/pasa.2017.65}, \href
  {https://ui.adsabs.harvard.edu/abs/2017PASA...34...69A} {34, e069}

\bibitem[\protect\citeauthoryear{{Astraatmadja} \&
  {Bailer-Jones}}{{Astraatmadja} \& {Bailer-Jones}}{2016}]{Astraatmadja2016}
{Astraatmadja} T.~L.,  {Bailer-Jones} C.~A.~L.,  2016, \mn@doi [\apj]
  {10.3847/1538-4357/833/1/119}, \href
  {http://adsabs.harvard.edu/abs/2016ApJ...833..119A} {833, 119}

\bibitem[\protect\citeauthoryear{{Bailer-Jones}}{{Bailer-Jones}}{2015}]{bailer-Jones2015}
{Bailer-Jones} C.~A.~L.,  2015, \mn@doi [\pasp] {10.1086/683116}, \href
  {http://adsabs.harvard.edu/abs/2015PASP..127..994B} {127, 994}

\bibitem[\protect\citeauthoryear{{Becker} et~al.,}{{Becker}
  et~al.}{2004}]{becker2004}
{Becker} A.~C.,  et~al., 2004, \mn@doi [\apj] {10.1086/421994}, \href
  {http://adsabs.harvard.edu/abs/2004ApJ...611..418B} {611, 418}

\bibitem[\protect\citeauthoryear{{Berger} et~al.,}{{Berger}
  et~al.}{2013}]{berger2013}
{Berger} E.,  et~al., 2013, \mn@doi [\apj] {10.1088/0004-637X/779/1/18}, \href
  {http://adsabs.harvard.edu/abs/2013ApJ...779...18B} {779, 18}

\bibitem[\protect\citeauthoryear{{Berthier}, {Vachier}, {Thuillot}, {Fernique},
  {Ochsenbein}, {Genova}, {Lainey}  \& {Arlot}}{{Berthier}
  et~al.}{2006}]{ber2006}
{Berthier} J.,  {Vachier} F.,  {Thuillot} W.,  {Fernique} P.,  {Ochsenbein} F.,
   {Genova} F.,  {Lainey} V.,   {Arlot} J.-E.,  2006, in {Gabriel} C.,
  {Arviset} C.,  {Ponz} D.,   {Enrique} S.,  eds,  Astronomical Society of the
  Pacific Conference Series Vol. 351, Astronomical Data Analysis Software and
  Systems XV. p.~367

\bibitem[\protect\citeauthoryear{{Bessell} \& {Murphy}}{{Bessell} \&
  {Murphy}}{2012}]{Bessell2012}
{Bessell} M.,  {Murphy} S.,  2012, \mn@doi [\pasp] {10.1086/664083}, \href
  {http://adsabs.harvard.edu/abs/2012PASP..124..140B} {124, 140}

\bibitem[\protect\citeauthoryear{{Bessell}, {Bloxham}, {Schmidt}, {Keller},
  {Tisserand}  \& {Francis}}{{Bessell} et~al.}{2011}]{Bessell2011}
{Bessell} M.,  {Bloxham} G.,  {Schmidt} B.,  {Keller} S.,  {Tisserand} P.,
  {Francis} P.,  2011, \mn@doi [\pasp] {10.1086/660849}, \href
  {http://adsabs.harvard.edu/abs/2011PASP..123..789B} {123, 789}

\bibitem[\protect\citeauthoryear{{Bochanski}, {West}, {Hawley}  \&
  {Covey}}{{Bochanski} et~al.}{2007}]{Bochanski2007}
{Bochanski} J.~J.,  {West} A.~A.,  {Hawley} S.~L.,   {Covey} K.~R.,  2007,
  \mn@doi [\aj] {10.1086/510240}, \href
  {http://adsabs.harvard.edu/abs/2007AJ....133..531B} {133, 531}

\bibitem[\protect\citeauthoryear{{Candelaresi}, {Hillier}, {Maehara},
  {Brandenburg}  \& {Shibata}}{{Candelaresi} et~al.}{2014}]{Candelaresi2014}
{Candelaresi} S.,  {Hillier} A.,  {Maehara} H.,  {Brandenburg} A.,   {Shibata}
  K.,  2014, \mn@doi [\apj] {10.1088/0004-637X/792/1/67}, \href
  {http://adsabs.harvard.edu/abs/2014ApJ...792...67C} {792, 67}

\bibitem[\protect\citeauthoryear{{Chang}, {Byun}  \& {Hartman}}{{Chang}
  et~al.}{2015}]{chang2015}
{Chang} S.-W.,  {Byun} Y.-I.,   {Hartman} J.~D.,  2015, \mn@doi [\apj]
  {10.1088/0004-637X/814/1/35}, \href
  {http://adsabs.harvard.edu/abs/2015ApJ...814...35C} {814, 35}

\bibitem[\protect\citeauthoryear{{Davenport}}{{Davenport}}{2016}]{Davenport2016}
{Davenport} J.~R.~A.,  2016, \mn@doi [\apj] {10.3847/0004-637X/829/1/23}, \href
  {http://adsabs.harvard.edu/abs/2016ApJ...829...23D} {829, 23}

\bibitem[\protect\citeauthoryear{{Davenport}, {Becker}, {Kowalski}, {Hawley},
  {Schmidt}, {Hilton}, {Sesar}  \& {Cutri}}{{Davenport}
  et~al.}{2012}]{davenport2012}
{Davenport} J.~R.~A.,  {Becker} A.~C.,  {Kowalski} A.~F.,  {Hawley} S.~L.,
  {Schmidt} S.~J.,  {Hilton} E.~J.,  {Sesar} B.,   {Cutri} R.,  2012, \mn@doi
  [\apj] {10.1088/0004-637X/748/1/58}, \href
  {http://adsabs.harvard.edu/abs/2012ApJ...748...58D} {748, 58}

\bibitem[\protect\citeauthoryear{{Frisch}, {Redfield}  \& {Slavin}}{{Frisch}
  et~al.}{2011}]{Frisch2011}
{Frisch} P.~C.,  {Redfield} S.,   {Slavin} J.~D.,  2011, \mn@doi [\araa]
  {10.1146/annurev-astro-081710-102613}, \href
  {https://ui.adsabs.harvard.edu/abs/2011ARA&A..49..237F} {49, 237}

\bibitem[\protect\citeauthoryear{{Gaia Collaboration} et~al.,}{{Gaia
  Collaboration} et~al.}{2016}]{gaia_2016}
{Gaia Collaboration} et~al., 2016, \mn@doi [\aap]
  {10.1051/0004-6361/201629272}, \href
  {http://adsabs.harvard.edu/abs/2016A%26A...595A...1G} {595, A1}

\bibitem[\protect\citeauthoryear{{Gaia Collaboration} et~al.,}{{Gaia
  Collaboration} et~al.}{2018a}]{gaia_dr2_2018}
{Gaia Collaboration} et~al., 2018a, \mn@doi [\aap]
  {10.1051/0004-6361/201833051}, \href
  {http://adsabs.harvard.edu/abs/2018A%26A...616A...1G} {616, A1}

\bibitem[\protect\citeauthoryear{{Gaia Collaboration} et~al.,}{{Gaia
  Collaboration} et~al.}{2018b}]{GaiaHDR2018}
{Gaia Collaboration} et~al., 2018b, \mn@doi [\aap]
  {10.1051/0004-6361/201832843}, \href
  {http://adsabs.harvard.edu/abs/2018A%26A...616A..10G} {616, A10}

\bibitem[\protect\citeauthoryear{{Gaia Collaboration} et~al.,}{{Gaia
  Collaboration} et~al.}{2018c}]{Gaia2018A&A...616A..11G}
{Gaia Collaboration} et~al., 2018c, \mn@doi [\aap]
  {10.1051/0004-6361/201832865}, \href
  {https://ui.adsabs.harvard.edu/abs/2018A&A...616A..11G} {616, A11}

\bibitem[\protect\citeauthoryear{{Gizis}}{{Gizis}}{2002}]{Gizis2002}
{Gizis} J.~E.,  2002, \mn@doi [\apj] {10.1086/341259}, \href
  {https://ui.adsabs.harvard.edu/abs/2002ApJ...575..484G} {575, 484}

\bibitem[\protect\citeauthoryear{{G{\"u}nther} et~al.,}{{G{\"u}nther}
  et~al.}{2019}]{Gunther2019arXiv}
{G{\"u}nther} M.~N.,  et~al., 2019, arXiv e-prints, \href
  {https://ui.adsabs.harvard.edu/abs/2019arXiv190100443G} {p. arXiv:1901.00443}

\bibitem[\protect\citeauthoryear{{Hawley} \& {Fisher}}{{Hawley} \&
  {Fisher}}{1992}]{Hawley1992}
{Hawley} S.~L.,  {Fisher} G.~H.,  1992, \mn@doi [\apjs] {10.1086/191640}, \href
  {http://adsabs.harvard.edu/abs/1992ApJS...78..565H} {78, 565}

\bibitem[\protect\citeauthoryear{{Hawley} et~al.,}{{Hawley}
  et~al.}{2003}]{Hawley2003}
{Hawley} S.~L.,  et~al., 2003, \mn@doi [\apj] {10.1086/378351}, \href
  {http://adsabs.harvard.edu/abs/2003ApJ...597..535H} {597, 535}

\bibitem[\protect\citeauthoryear{{Hilton}, {West}, {Hawley}  \&
  {Kowalski}}{{Hilton} et~al.}{2010}]{hilton2010}
{Hilton} E.~J.,  {West} A.~A.,  {Hawley} S.~L.,   {Kowalski} A.~F.,  2010,
  \mn@doi [\aj] {10.1088/0004-6256/140/5/1402}, \href
  {http://adsabs.harvard.edu/abs/2010AJ....140.1402H} {140, 1402}

\bibitem[\protect\citeauthoryear{{Ho} et~al.,}{{Ho} et~al.}{2018}]{ho2018}
{Ho} A.~Y.~Q.,  et~al., 2018, \mn@doi [\apjl] {10.3847/2041-8213/aaaa62}, \href
  {http://adsabs.harvard.edu/abs/2018ApJ...854L..13H} {854, L13}

\bibitem[\protect\citeauthoryear{{Howard} et~al.,}{{Howard}
  et~al.}{2018}]{howard2018}
{Howard} W.~S.,  et~al., 2018, \mn@doi [\apjl] {10.3847/2041-8213/aacaf3},
  \href {http://adsabs.harvard.edu/abs/2018ApJ...860L..30H} {860, L30}

\bibitem[\protect\citeauthoryear{{Kowalski}, {Hawley}, {Hilton}, {Becker},
  {West}, {Bochanski}  \& {Sesar}}{{Kowalski} et~al.}{2009}]{kow2009}
{Kowalski} A.~F.,  {Hawley} S.~L.,  {Hilton} E.~J.,  {Becker} A.~C.,  {West}
  A.~A.,  {Bochanski} J.~J.,   {Sesar} B.,  2009, \mn@doi [\aj]
  {10.1088/0004-6256/138/2/633}, \href
  {http://adsabs.harvard.edu/abs/2009AJ....138..633K} {138, 633}

\bibitem[\protect\citeauthoryear{{Kowalski}, {Hawley}, {Wisniewski}, {Osten},
  {Hilton}, {Holtzman}, {Schmidt}  \& {Davenport}}{{Kowalski}
  et~al.}{2013}]{kow2013}
{Kowalski} A.~F.,  {Hawley} S.~L.,  {Wisniewski} J.~P.,  {Osten} R.~A.,
  {Hilton} E.~J.,  {Holtzman} J.~A.,  {Schmidt} S.~J.,   {Davenport} J.~R.~A.,
  2013, \mn@doi [\apjs] {10.1088/0067-0049/207/1/15}, \href
  {http://adsabs.harvard.edu/abs/2013ApJS..207...15K} {207, 15}

\bibitem[\protect\citeauthoryear{{Kowalski} et~al.,}{{Kowalski}
  et~al.}{2019}]{Kowalski2019}
{Kowalski} A.~F.,  et~al., 2019, \mn@doi [\apj] {10.3847/1538-4357/aaf058},
  \href {http://adsabs.harvard.edu/abs/2019ApJ...871..167K} {871, 167}

\bibitem[\protect\citeauthoryear{{Kulkarni} \& {Rau}}{{Kulkarni} \&
  {Rau}}{2006}]{Kulkarni2006}
{Kulkarni} S.~R.,  {Rau} A.,  2006, \mn@doi [\apj] {10.1086/505423}, \href
  {https://ui.adsabs.harvard.edu/abs/2006ApJ...644L..63K} {644, L63}

\bibitem[\protect\citeauthoryear{{Luri} et~al.,}{{Luri}
  et~al.}{2014}]{Luri2014}
{Luri} X.,  et~al., 2014, \mn@doi [\aap] {10.1051/0004-6361/201423636}, \href
  {https://ui.adsabs.harvard.edu/abs/2014A&A...566A.119L} {566, A119}

\bibitem[\protect\citeauthoryear{{Luri} et~al.,}{{Luri}
  et~al.}{2018}]{Luri2018}
{Luri} X.,  et~al., 2018, \mn@doi [\aap] {10.1051/0004-6361/201832964}, \href
  {http://adsabs.harvard.edu/abs/2018A%26A...616A...9L} {616, A9}

\bibitem[\protect\citeauthoryear{{MacGregor}, {Weinberger}, {Wilner},
  {Kowalski}  \& {Cranmer}}{{MacGregor} et~al.}{2018}]{MacGregor2018ApJ}
{MacGregor} M.~A.,  {Weinberger} A.~J.,  {Wilner} D.~J.,  {Kowalski} A.~F.,
  {Cranmer} S.~R.,  2018, \mn@doi [\apjl] {10.3847/2041-8213/aaad6b}, \href
  {http://adsabs.harvard.edu/abs/2018ApJ...855L...2M} {855, L2}

\bibitem[\protect\citeauthoryear{{Maehara} et~al.,}{{Maehara}
  et~al.}{2012}]{Maehara2012}
{Maehara} H.,  et~al., 2012, \mn@doi [\nat] {10.1038/nature11063}, \href
  {http://adsabs.harvard.edu/abs/2012Natur.485..478M} {485, 478}

\bibitem[\protect\citeauthoryear{{Miller} et~al.,}{{Miller}
  et~al.}{2001}]{Miller2001}
{Miller} C.~J.,  et~al., 2001, \mn@doi [\aj] {10.1086/324109}, \href
  {http://adsabs.harvard.edu/abs/2001AJ....122.3492M} {122, 3492}

\bibitem[\protect\citeauthoryear{{Mowlavi} et~al.,}{{Mowlavi}
  et~al.}{2018}]{mowlavi2018}
{Mowlavi} N.,  et~al., 2018, \mn@doi [\aap] {10.1051/0004-6361/201833366},
  \href {https://ui.adsabs.harvard.edu/abs/2018A&A...618A..58M} {618, A58}

\bibitem[\protect\citeauthoryear{{Notsu} et~al.,}{{Notsu}
  et~al.}{2013}]{Notsu2013}
{Notsu} Y.,  et~al., 2013, \mn@doi [\apj] {10.1088/0004-637X/771/2/127}, \href
  {http://adsabs.harvard.edu/abs/2013ApJ...771..127N} {771, 127}

\bibitem[\protect\citeauthoryear{Onken et~al.,}{Onken et~al.}{2019}]{Onken2019}
Onken C.~A.,  et~al., 2019, \mn@doi [Publications of the Astronomical Society
  of Australia] {10.1017/pasa.2019.27}, 36, e033

\bibitem[\protect\citeauthoryear{{Paudel}, {Gizis}, {Mullan}, {Schmidt},
  {Burgasser}, {Williams}  \& {Berger}}{{Paudel}
  et~al.}{2018a}]{Paudel2018ApJ...858...55P}
{Paudel} R.~R.,  {Gizis} J.~E.,  {Mullan} D.~J.,  {Schmidt} S.~J.,  {Burgasser}
  A.~J.,  {Williams} P.~K.~G.,   {Berger} E.,  2018a, \mn@doi [\apj]
  {10.3847/1538-4357/aab8fe}, \href
  {http://adsabs.harvard.edu/abs/2018ApJ...858...55P} {858, 55}

\bibitem[\protect\citeauthoryear{{Paudel}, {Gizis}, {Mullan}, {Schmidt},
  {Burgasser}, {Williams}  \& {Berger}}{{Paudel}
  et~al.}{2018b}]{Paudel2018ApJ...861...76P}
{Paudel} R.~R.,  {Gizis} J.~E.,  {Mullan} D.~J.,  {Schmidt} S.~J.,  {Burgasser}
  A.~J.,  {Williams} P.~K.~G.,   {Berger} E.,  2018b, \mn@doi [\apj]
  {10.3847/1538-4357/aac8e0}, \href
  {http://adsabs.harvard.edu/abs/2018ApJ...861...76P} {861, 76}

\bibitem[\protect\citeauthoryear{{Pickles}}{{Pickles}}{1998}]{Pickles1998}
{Pickles} A.~J.,  1998, \mn@doi [\pasp] {10.1086/316197}, \href
  {https://ui.adsabs.harvard.edu/abs/1998PASP..110..863P} {110, 863}

\bibitem[\protect\citeauthoryear{{Rabus} et~al.,}{{Rabus}
  et~al.}{2019}]{Rabus2019}
{Rabus} M.,  et~al., 2019, \mn@doi [\mnras] {10.1093/mnras/sty3430}, \href
  {https://ui.adsabs.harvard.edu/abs/2019MNRAS.484.2674R} {484, 2674}

\bibitem[\protect\citeauthoryear{{Robin} et~al.,}{{Robin}
  et~al.}{2012}]{Robin2012}
{Robin} A.~C.,  et~al., 2012, \mn@doi [\aap] {10.1051/0004-6361/201118646},
  \href {https://ui.adsabs.harvard.edu/abs/2012A&A...543A.100R} {543, A100}

\bibitem[\protect\citeauthoryear{{Schlegel}, {Finkbeiner}  \&
  {Davis}}{{Schlegel} et~al.}{1998}]{Schlegel1998}
{Schlegel} D.~J.,  {Finkbeiner} D.~P.,   {Davis} M.,  1998, \mn@doi [\apj]
  {10.1086/305772}, \href
  {https://ui.adsabs.harvard.edu/abs/1998ApJ...500..525S} {500, 525}

\bibitem[\protect\citeauthoryear{{Schmidt} et~al.,}{{Schmidt}
  et~al.}{2014}]{Schmidt2014}
{Schmidt} S.~J.,  et~al., 2014, \mn@doi [\apjl] {10.1088/2041-8205/781/2/L24},
  \href {http://adsabs.harvard.edu/abs/2014ApJ...781L..24S} {781, L24}

\bibitem[\protect\citeauthoryear{{Schmidt} et~al.,}{{Schmidt}
  et~al.}{2016}]{Schmidt2016}
{Schmidt} S.~J.,  et~al., 2016, \mn@doi [\apjl] {10.3847/2041-8205/828/2/L22},
  \href {http://adsabs.harvard.edu/abs/2016ApJ...828L..22S} {828, L22}

\bibitem[\protect\citeauthoryear{{Schmidt} et~al.,}{{Schmidt}
  et~al.}{2019}]{Schmidt2019}
{Schmidt} S.~J.,  et~al., 2019, \mn@doi [\apj] {10.3847/1538-4357/ab148d},
  \href {https://ui.adsabs.harvard.edu/abs/2019ApJ...876..115S} {876, 115}

\bibitem[\protect\citeauthoryear{{Shappee} et~al.,}{{Shappee}
  et~al.}{2017}]{Shappee2017}
{Shappee} B.~J.,  et~al., 2017, \mn@doi [Science] {10.1126/science.aaq0186},
  \href {https://ui.adsabs.harvard.edu/abs/2017Sci...358.1574S} {358, 1574}

\bibitem[\protect\citeauthoryear{{Soraisam}, {Gilfanov}, {Kupfer}, {Prince},
  {Masci}, {Laher}  \& {Kong}}{{Soraisam} et~al.}{2018}]{Soraisam2018}
{Soraisam} M.~D.,  {Gilfanov} M.,  {Kupfer} T.,  {Prince} T.~A.,  {Masci} F.,
  {Laher} R.~R.,   {Kong} A.~K.~H.,  2018, \mn@doi [\aap]
  {10.1051/0004-6361/201732068}, \href
  {http://adsabs.harvard.edu/abs/2018A%26A...615A.152S} {615, A152}

\bibitem[\protect\citeauthoryear{{Stassun} et~al.,}{{Stassun}
  et~al.}{2011}]{Stassun2011}
{Stassun} K.~G.,  et~al., 2011, in {Johns-Krull} C.,  {Browning} M.~K.,
  {West} A.~A.,  eds,  Astronomical Society of the Pacific Conference Series
  Vol. 448, 16th Cambridge Workshop on Cool Stars, Stellar Systems, and the
  Sun. p.~505 (\mn@eprint {arXiv} {1012.2580})

\bibitem[\protect\citeauthoryear{{Walkowicz} et~al.,}{{Walkowicz}
  et~al.}{2011}]{Walkowicz2011}
{Walkowicz} L.~M.,  et~al., 2011, \mn@doi [\aj] {10.1088/0004-6256/141/2/50},
  \href {https://ui.adsabs.harvard.edu/abs/2011AJ....141...50W} {141, 50}

\bibitem[\protect\citeauthoryear{{Watson}, {Henden}  \& {Price}}{{Watson}
  et~al.}{2006}]{Watson2006}
{Watson} C.~L.,  {Henden} A.~A.,   {Price} A.,  2006, Society for Astronomical
  Sciences Annual Symposium, \href
  {https://ui.adsabs.harvard.edu/abs/2006SASS...25...47W} {25, 47}

\bibitem[\protect\citeauthoryear{{Watson}, {Henden}  \& {Price}}{{Watson}
  et~al.}{2017}]{Watson2017}
{Watson} C.,  {Henden} A.~A.,   {Price} A.,  2017, VizieR Online Data Catalog,
  \href {https://ui.adsabs.harvard.edu/abs/2017yCat....102027W} {p. B/vsx}

\bibitem[\protect\citeauthoryear{{West}, {Bochanski}, {Hawley}, {Cruz},
  {Covey}, {Silvestri}, {Reid}  \& {Liebert}}{{West} et~al.}{2006}]{west2006}
{West} A.~A.,  {Bochanski} J.~J.,  {Hawley} S.~L.,  {Cruz} K.~L.,  {Covey}
  K.~R.,  {Silvestri} N.~M.,  {Reid} I.~N.,   {Liebert} J.,  2006, \mn@doi
  [\aj] {10.1086/508652}, \href
  {https://ui.adsabs.harvard.edu/abs/2006AJ....132.2507W} {132, 2507}

\bibitem[\protect\citeauthoryear{{West}, {Hawley}, {Bochanski}, {Covey},
  {Reid}, {Dhital}, {Hilton}  \& {Masuda}}{{West} et~al.}{2008}]{West2008}
{West} A.~A.,  {Hawley} S.~L.,  {Bochanski} J.~J.,  {Covey} K.~R.,  {Reid}
  I.~N.,  {Dhital} S.,  {Hilton} E.~J.,   {Masuda} M.,  2008, \mn@doi [\aj]
  {10.1088/0004-6256/135/3/785}, \href
  {https://ui.adsabs.harvard.edu/abs/2008AJ....135..785W} {135, 785}

\bibitem[\protect\citeauthoryear{{West} et~al.,}{{West}
  et~al.}{2011}]{west2011}
{West} A.~A.,  et~al., 2011, \mn@doi [\aj] {10.1088/0004-6256/141/3/97}, \href
  {http://adsabs.harvard.edu/abs/2011AJ....141...97W} {141, 97}

\bibitem[\protect\citeauthoryear{{Winters} et~al.,}{{Winters}
  et~al.}{2015}]{winters2015}
{Winters} J.~G.,  et~al., 2015, \mn@doi [\aj] {10.1088/0004-6256/149/1/5},
  \href {http://adsabs.harvard.edu/abs/2015AJ....149....5W} {149, 5}

\bibitem[\protect\citeauthoryear{{Wolf} et~al.,}{{Wolf}
  et~al.}{2018}]{wolf2018}
{Wolf} C.,  et~al., 2018, \mn@doi [\pasa] {10.1017/pasa.2018.5}, \href
  {http://adsabs.harvard.edu/abs/2018PASA...35...10W} {35, e010}

\bibitem[\protect\citeauthoryear{{York} et~al.,}{{York}
  et~al.}{2000}]{York2000}
{York} D.~G.,  et~al., 2000, \mn@doi [\aj] {10.1086/301513}, \href
  {https://ui.adsabs.harvard.edu/abs/2000AJ....120.1579Y} {120, 1579}

\bibitem[\protect\citeauthoryear{{van Roestel} et~al.,}{{van Roestel}
  et~al.}{2019}]{vanRoestel2019}
{van Roestel} J.,  et~al., 2019, \mn@doi [\mnras] {10.1093/mnras/stz241}, \href
  {http://adsabs.harvard.edu/abs/2019MNRAS.484.4507V} {484, 4507}

\makeatother
\end{thebibliography}

\end{document}